\documentclass[lettersize,journal]{IEEEtran}
\usepackage{amsmath,amsfonts}
\usepackage{array}
\usepackage[caption=false,font=normalsize,labelfont=sf,textfont=sf]{subfig}
\usepackage{textcomp}
\usepackage{stfloats}
\usepackage{url}
\usepackage{verbatim}
\usepackage{graphicx}
\usepackage{upgreek}
\usepackage{color, soul}
\usepackage{cite}
\usepackage{xurl}
\usepackage[hidelinks,breaklinks=true]{hyperref}
\begin{document}

\title{Channel-Aware Behavioral Power Modeling of CMOS OOK Transceivers for Wireless Network-on-Chip Systems}

\author{Mohammad Shahmoradi,~\IEEEmembership{Student Member, IEEE,}
Ahmet Yelboğa,~\IEEEmembership{Student Member, IEEE,}
Eduard Alarcón,~\IEEEmembership{Fellow, IEEE,}
Korkut Kaan Tokgöz,~\IEEEmembership{Member, IEEE,}
and Sergi Abadal,~\IEEEmembership{Senior Member, IEEE}%
\thanks{Mohammad Shahmoradi, Eduard Alarcón, and Sergi Abadal are with the Universitat Politècnica de Catalunya (UPC), Barcelona, Spain (e-mail: mohammad.shahmoradi@upc.edu). Ahmet Yelboğa and Korkut Kaan Tokgöz are with Sabancı University, Istanbul, Türkiye.}

}

\markboth{IEEE Transactions on Circuits and Systems---I: Regular Papers}%
{Shahmoradi \MakeLowercase{\textit{et al.}}: Channel-Aware Behavioral Power Modeling of CMOS OOK Transceivers for WNoC}

\maketitle

\begin{abstract}
Wireless Network-on-Chip (WNoC) systems enable low-latency communication in many-core platforms through short-range wireless links. However, the power consumption of integrated transceivers (TRXs), dominated by that of the RF front-end circuitry, remains a major challenge. Moreover, the optimal operating frequency is still unclear, as bandwidth, energy efficiency, and technology maturity must be balanced. This work presents a channel-aware behavioral modeling framework to estimate power consumption and identify energy-efficient operating points in non-coherent On-Off Keying (OOK) TRXs over a wide frequency range. The approach leverages survey data from CMOS implementations to derive frequency-dependent power models for key TRX sub-blocks, including the power amplifier (PA), oscillator, mixer, low noise amplifier (LNA), and envelope detector (ED). By incorporating the frequency-dependent channel loss into the TRX power budget, the model captures system-level power trade-offs across operating regimes. The analysis reveals a frequency-dependent shift in power dominance between the transmitter and receiver: oscillator- and ED-dominated regimes at lower frequencies transition to PA- and LNA-dominated behavior at higher frequencies. Furthermore, the energy-per-bit landscape exhibits sweet spots and a model-based global minimum, indicating that optimal operation cannot be achieved by optimizing transmitter or receiver independently. Overall, the proposed framework enables rapid and physically grounded exploration of power scaling with frequency and channel conditions, providing practical guidelines for energy-efficient design of high-frequency wireless links for WNoC systems and beyond.
\end{abstract}

\vspace{-0.3cm}
\section{Introduction}
\IEEEPARstart{T}{he} number of processing cores integrated on a single chip has increased steadily to meet the growing demand for computational performance. However, this scaling trend has exposed fundamental limitations in conventional wired on-chip interconnects \cite{das2024chip}. As the amount of data exchanged among cores continues to grow, metal-based interconnects struggle to sustain bandwidth scalability, low latency, and energy efficiency. These constraints limit the performance gains expected from higher integration levels and motivate the exploration of alternative communication paradigms. Among the proposed solutions, Wireless Network-on-Chip (WNoC) architectures have emerged as a potential approach to enable scalable, low-latency, and broadcast-capable communication across distant cores~\cite{tapie2024systematic, bandara2026towards, wang2014wireless, gaha2022novel, weerasena2024security, abadal2018medium}.

Despite the conceptual advantages of the WNoC paradigm, the practical implementation of such systems remains challenging. Integrating multiple wireless transceivers (TRXs) on a chip imposes stringent constraints on both silicon area and energy consumption, since each TRX must operate within the limited power budget of the overall system. Power consumption plays a critical role, as it directly affects not only the energy efficiency of the wireless link but also the processor's thermal stability and long-term reliability. Consequently, accurately understanding and managing the power consumption of the TRX RF front end is a key requirement for the realization of practical WNoC architectures.

\begin{figure}[t]
\centering
\includegraphics[width=90mm]{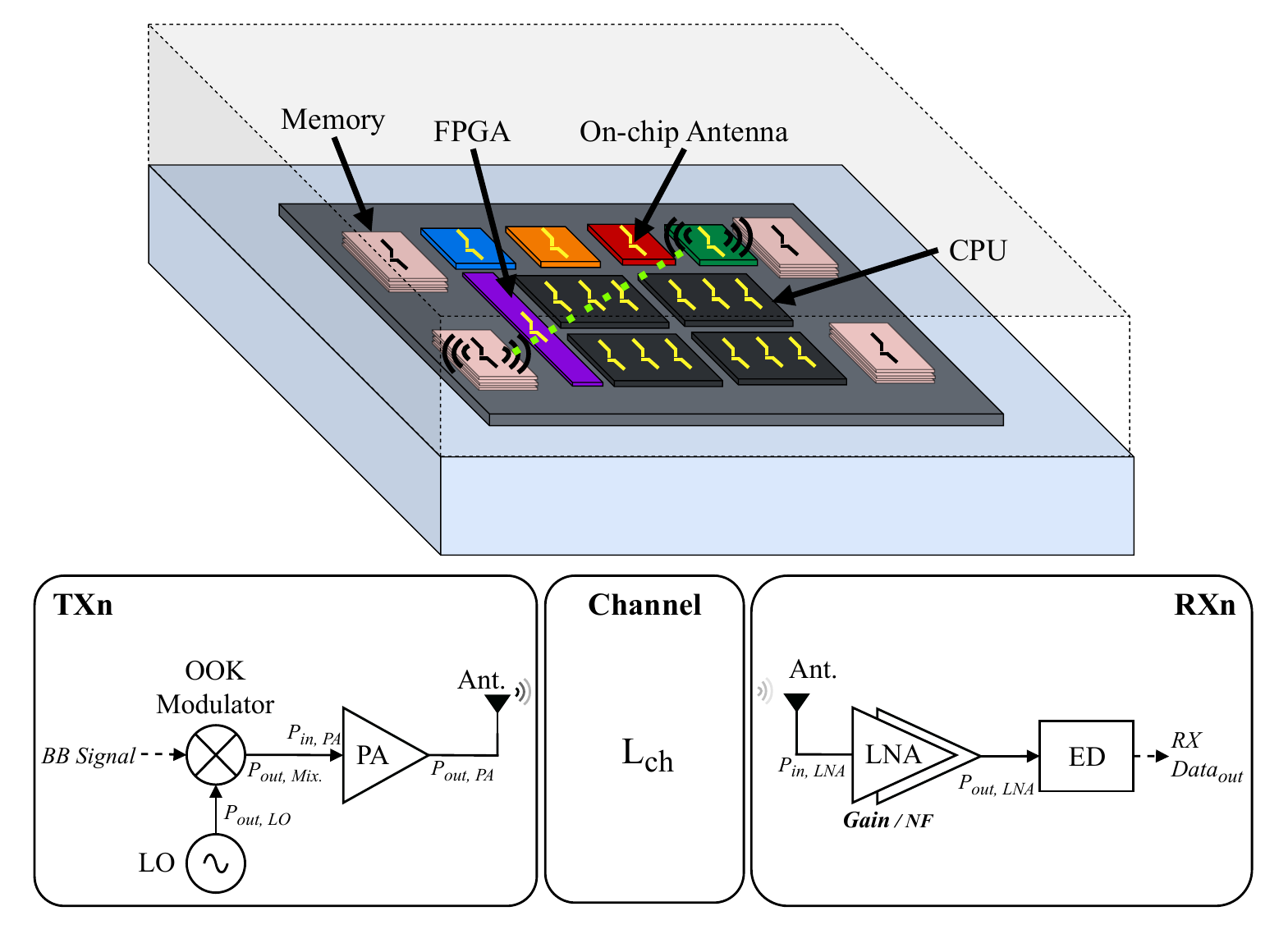}
\vspace{-0.4cm}
\caption{Representative WNoC architecture (top) and the corresponding abstraction of the TRX RF analog front-end and channel path-loss model used in this work (bottom).}\label{fig:WNoC_architecture_sub_block}
\vspace{-0.2cm}
\end{figure}

In addition, the optimal operating frequency for WNoC systems remains an open question. While higher frequencies in the millimeter-wave (mm-wave) and terahertz (THz) bands provide wider bandwidth and enable smaller antenna dimensions, they may also incur higher propagation losses, stronger circuit non-idealities, and decreased efficiency. A quantitative understanding of how TRX power consumption scales with frequency is therefore essential to guide early design decisions and to enable energy-efficient WNoC implementations.

Different TRX architectures and circuit topologies have been explored to enable on-chip wireless communication. Common modulation schemes include Amplitude Shift Keying (ASK), Quadrature Phase-Shift Keying (QPSK), and On-Off Keying (OOK)~\cite{chen20096, marcu200990, byeon201367, lee2010low, juntunen201060, byeon20202, peng202426}. Among these modulation schemes, OOK is frequently preferred for WNoC links, due to the possibility of using non-coherent detection, which leads to simple circuits and low power requirements. For example, fully integrated OOK transmitters operating at 60 GHz have demonstrated multi-gigabit data rates with power consumption of several tens of milliwatts~\cite{byeon201367}. This range is consistent with the order of magnitude predicted by the system-level analysis presented in this work. These examples highlight that modulation choice and circuit architecture strongly influence the power-performance tradeoff in WNoC systems. For clearer insight, Fig.~\ref{fig:WNoC_architecture_sub_block} depicts a representative WNoC architecture along with the associated abstraction of the TRX analog front-end and the propagation channel.

Given that full prototype implementations of WNoC TRXs are both time-consuming and costly, early design exploration often relies on modeling approaches derived from surveys of reported circuit implementations. Several surveys and studies have investigated different aspects of on-chip wireless communication, including wireless channel characterization, on-chip antenna design, and communication protocols relevant to WNoC architectures~\cite{abadal2019wave, gade2019millimeter, deb2012wireless, lemic2021survey, ganguly2018advances}, leaving the detailed power behavior of the RF transceiver as an open issue. 
 

To address this gap, this work leverages survey data (including our own surveys) from reported CMOS implementations of key RF building blocks, including power amplifiers (PAs)~\cite{wang2020power}, oscillators~\cite{shahmoradi2025vcosurvey}, mixers~\cite{shahmoradi2025mixersurvey}, and low-noise amplifiers (LNAs)~\cite{belostotski2020low}. Although many of these circuits were originally developed for other wireless applications such as contactless links or data kiosks, they provide valuable insight into the performance trends of each RF front-end component as a function of the operating frequency. 

In this context, this paper develops a behavioral model that, for the first time, quantifies the power consumption of the entire RF front-end of a non-coherent OOK transceiver over a broad frequency range from 28 to 245 GHz and on a sub-block basis. 
By modeling the required sub-blocks within a unified framework and relating the power consumption to the signal levels across the RF chains, the model enables a systematic evaluation of how WNoC energy efficiency evolves in different operating conditions, thereby supporting informed architectural and technology design choices.

This work builds on our previous research on transmitter power consumption~\cite{shahmoradi2025frequency}, which has been expanded very significantly by introducing our own mixer and oscillator datasets \cite{shahmoradi2025mixersurvey, shahmoradi2025vcosurvey} at the transmitter, adding the entire receiver model, and exploring the transmitter-receiver tradeoffs. The work differs significantly from other modeling efforts from the literature, which either focus on single sub-blocks \cite{wu2023optimized}, model the transceiver at a single specific band for applications vastly different to WNoC \cite{hietanen2025technology, paoli2025fast}, or do not model the transceiver block by block and hence miss the capability to navigate the design tradeoffs in a frequency-dependent manner \cite{abadal2014area}.  


The remainder of this paper is organized as follows. Section~\ref{Section: Methodology} presents the overall system model, while Sections~\ref{Section: Methodology_TX} and~\ref{Section: Methodology_RX} describe the modeling strategies adopted for the transmitter and receiver sub-blocks, respectively. Section~\ref{Section: Discussion} combines the sub-block models to analyze the behavior of the complete transceiver, whereas Section~\ref{Section: Discussion2} discusses the resulting system-level implications. Finally, Section~\ref{Section: Conclusion} concludes the paper.

\vspace{-0.2cm}
\section{System Model Overview} \label{Section: Methodology}

At the system level, the end-to-end wireless link in a WNoC architecture can be modeled as three interrelated abstract blocks: the transmitter, the propagation channel, and the receiver, as shown in Fig.~\ref{fig:WNoC_architecture_sub_block}. In the proposed framework, the transmitter block aggregates the DC power consumption of the PA, local oscillator (e.g., voltage-controlled oscillator (VCO)), and mixer, while the receiver block accounts for the LNA and ED. Finally, the total power consumption of TRX is expressed as $P_{\mathrm{DC,TRX}} = P_{\mathrm{DC,TX}} + P_{\mathrm{DC,RX}}$. 
\vspace{-0.2cm}
\subsection{Link Budget}
\label{subsec:link_budget}

To ensure reliable data transmission, the system must satisfy the link power budget constraints, whereby the target bit error rate (BER) determines the required transmitted power through receiver sensitivity requirements and propagation losses.

For OOK modulation with envelope detection over an Additive White Gaussian Noise (AWGN) channel, the BER can be approximated with
\begin{equation}
\mathrm{BER} = \tfrac{1}{2}\exp\!\left(-\tfrac{E_b}{2N_0}\right).
\label{Eqn:BER}
\end{equation}

This relationship defines the required ratio of the received energy per bit, $E_b$, to the spectral density of the noise power, $N_0$, to satisfy a given BER constraint. In particular, achieving $\mathrm{BER}=10^{-12}$ requires approximately $(E_b/N_0)\approx17.5$ dB.

To relate this requirement to receiver performance, note that $E_b = P_{\mathrm{r}}/R_b$, $N_0 = P_N/B$, and $\mathrm{SNR} = P_{\mathrm{r}}/P_N$ where $R_b$ is the data rate and $B$ is the receiver equivalent noise bandwidth (ENBW) at the detector input. Accordingly, the bit energy-to-noise power spectral density ratio can be expressed as
\begin{equation}
\frac{E_b}{N_0} = \frac{P_r}{R_b N_0}.
\label{Eqn:Eb_N0}
\end{equation}

For OOK modulation, the signal spectrum extends approximately up to $2R_b$ in the main lobe or $R_b$ in one-sided bandwidth. Since the receiver must accommodate this bandwidth, the ENBW is assumed to be of the same order, leading to the approximation $B \approx R_b$~\cite{yelboga2025dynamic}.

From the considerations above, for a fixed BER target (i.e., fixed $E_b/N_0$), increasing the bit rate leads to a proportional increase in the receiver bandwidth and hence the integrated noise power. As a result, the SNR decreases unless the received signal power is increased. 

From another perspective, the total receiver noise power, referred to the receiver input, can be expressed as
\begin{equation}
P_N = k T_0 F B,
\end{equation}
where \(k\) is Boltzmann’s constant, \(T_0=290\) K is the reference temperature, and \(F\) is the receiver noise factor. Substituting this relation into Eqn.~\eqref{Eqn:Eb_N0} gives
\begin{equation}
\frac{E_b}{N_0}
=
\frac{P_{\mathrm{r}}}{k T_0 F R_b}.
\label{Eqn:NF_With_EB_N0}
\end{equation}

Therefore, for a fixed received power, increasing the bit rate requires a lower receiver noise factor, i.e., a lower noise figure (NF), to maintain the same BER. This assumption is adopted in this work to model the energy-per-bit behavior under fixed received power conditions.

To define the minimum required signal power, the receiver sensitivity $P_{\mathrm{sen}}$ is given by
\begin{equation}
P_{\mathrm{sen}} = -174 + NF + 10\log_{10}B + \mathrm{SNR}_{\min},
\label{eq:sensitivity_rewritten}
\end{equation}
where all quantities are in dB. According to Friis’ cascaded noise model~\cite{razavi2012rf}, the overall receiver NF is typically dominated by the first amplification stage and can be approximated by that of the LNA.

\begin{table}[t]
\centering 
\caption{Representative high-speed CMOS receiver sensitivity with
corresponding power consumption and other specs.}
\vspace{-0.2cm}
\label{tab:rx_sens} 
\renewcommand{\arraystretch}{1.15} 
\setlength{\tabcolsep}{4pt}
\footnotesize
\begin{tabular}{c c c c c c}
\hline
Ref. & $f_{c}$ (GHz) & $R_{\mathrm{b}}$ (Gb/s) & $P_{\mathrm{sens}}$ (dBm) & $P_{\mathrm{DC,RX}}$ (mW) & BER \\
\hline
\cite{byeon20202} & 60 & 12.5 & $-25$ & 21 & $10^{-12}$ \\
\cite{byeon201367} & 60 & 10.7 & $-32.5$ & 67 & $10^{-12}$ \\
\cite{lee201620} & 84 & 20 & $-29$ & 46 & $10^{-12}$ \\
\cite{lee201620} & 84 & 10 & $-43$ & 46 & $10^{-12}$ \\
\cite{katayama2013209mw} & 134 & 10 & $-33.7$ & 132 & $10^{-12}$ \\
\cite{morath2024designing} & 60 & 2.8 & $-20.5$ & 15.3 & $10^{-11}$ \\
\hline
\vspace{-0.75cm}
\end{tabular}
\end{table}

To establish realistic modeling assumptions, representative CMOS receivers reported in the literature are summarized in Table~\ref{tab:rx_sens}. Based on these results, a sensitivity range of $-25$ to $-43$ dBm is considered adequate for high-speed mmWave receivers and is used in subsequent analysis. Also, according to Eqns.~\eqref{Eqn:BER} and \eqref{Eqn:Eb_N0}, for a target BER of $10^{-12}$, the minimum SNR required is approximately $17.5$ dB.

Once $\mathrm{SNR}_{\min}$ and $P_{\mathrm{sen}}$ are fixed, the allowable receiver NF becomes dependent on the bit rate. For representative bit rates of $\{2.2, 6, 11, 20\}$ Gbps, the corresponding maximum allowable NF values (assuming $P_{\mathrm{sen}}=-35$ dBm) are approximately $\{28.1, 23.7, 21.1, 18.5\}$ dB. Therefore, to satisfy the link budget, the receiver NF must remain below these values. In fact, as it will be observed later in the paper, we use this relationship to explore the efficiency of transceivers when tightening or relaxing the NF requirements on the receiver.

\vspace{-0.2cm}
\subsection{Channel Considerations}
\label{sec:channel}
The link budget relates the transmitted power, antenna gains, and channel loss. The received power $P_{\mathrm{r}}$, which must satisfy $P_{\mathrm{r}} \ge P_{\mathrm{sens}}$, is expressed as
\begin{equation}
    P_{\mathrm{r}} = P_{\mathrm{t}} + G_{\mathrm{t}} + G_{\mathrm{r}} - L_{\mathrm{ch}},
    \label{Eqn:Ferris}
\end{equation}
where $P_{\mathrm{r}}$ and $P_{\mathrm{t}}$ are the received and transmitter powers, $G_{\mathrm{t}}$ and $G_{\mathrm{r}}$ are the gains of the transmitter and receiver antennas, and $L_{\mathrm{ch}}$ is the path loss of the channel, all in dB units. In this work, antennas are not modeled as power-consuming components and are assumed to be omnidirectional with 0 dB gain. Under this assumption, Eqn.~\eqref{Eqn:Ferris} simplifies to
\begin{equation}
    P_{\mathrm{r}} = P_{\mathrm{t}} - L_{\mathrm{ch}}.
    \label{Eqn:PRX_PTX_LCH}
\end{equation}

Wireless propagation within a chip package differs fundamentally from free-space communication due to dielectric losses in silicon and oxide layers, reflections, and multipath effects caused by package boundaries and metal structures~\cite{agyeman2016efficient,el2018accurate}. As a result, the channel strongly depends on both frequency and packaging configuration. This behavior has been extensively characterized through full-wave electromagnetic simulations and measurements in prior works~\cite{abadal2019wave,timoneda2018channel,gade2018chip}.

For a fixed package geometry, the path loss in dB can be described using the log-distance model
\begin{equation}
L_{\mathrm{ch}}(f,d)=A(f)+10n\log_{10}\!\left(\frac{d}{d_0}\right),
\label{eq:path_loss_model}
\end{equation}
where $A(f)$ is a frequency-dependent intercept loss and $n$ is the path-loss exponent. Reported values of $n$ in optimized flip-chip environments lie in the range of $0.75$-$1.4$~\cite{abadal2019wave}, indicating that the path loss exponent is primarily determined by the cavity geometry rather than the carrier frequency. Accordingly, $n$ is treated as frequency-independent, while the frequency dependence of the channel is captured through $A(f)$.

To enable system-level analysis, a simplified channel abstraction is adopted. Instead of relying on a specific channel model, representative path loss values are used as input parameters. This is motivated by the fact that channel attenuation in WNoC systems strongly depends on several factors, including antenna design, chip layout, and packaging configuration, and therefore does not follow a unique general trend.

Accordingly, following results from the literature \cite{abadal2019wave}, $L_{ch}$ is modeled as $\{24, 28, 32, 36\}$~dB for operating frequencies of $\{28, 60, 140, 245\}$~GHz, respectively, assuming a TX-RX separation of 5~mm. These values are selected to represent a moderate-loss CMOS-based scenario and to enable a consistent comparison across different frequency points.

\begin{table*}[t]
\centering
\caption{Overview of the system-level modeled TRX RF front-end sub-blocks.}
\vspace{-0.2cm}
\label{tab:all_behavioral_refs}
\begin{tabular}{ p{1.7cm} p{2.3cm} p{5.3cm} p{1.5cm}}
\hline
\textbf{Sub-block} & \textbf{Method} & \textbf{Open Parameters} & \textbf{Reference} \\
\hline

PA    
& Survey-based 
& Frequency, $P_{\mathrm{out}}$, efficiency
&\cite{wang2020power}\\

VCO   
& Survey-based
& Frequency, $P_{\mathrm{RF,out}}$, DC-to-RF efficiency
&\cite{wang2020power, shahmoradi2025vcosurvey}\\

Mixer 
& Survey-based
& Frequency, conversion gain
&\cite{shahmoradi2025mixersurvey}\\
\hline

LNA   
& Survey-based 
& Frequency, gain, noise factor (F)
&\cite{belostotski2020low}\\

ED    
& Simulation-based 
& Frequency, RF input power
& N/A\\

\hline

\end{tabular}
\vspace{-0.3cm}
\end{table*}

\vspace{-0.2cm}

\subsection{Transceiver Model Overview}
The proposed power model is formulated by modeling each sub-block of the non-coherent OOK TRX independently at the system level. Table~\ref{tab:all_behavioral_refs} summarizes the approach taken for each sub-block, together with the parameters left open for exploration. The transmitter side power modeling and the receiver-side sensitivity-driven constraints are detailed in Sections~\ref{Section: Methodology_TX} and~\ref{Section: Methodology_RX}, respectively.

Following this approach, the total power consumption of the transmitter and receiver chains can be expressed as
\begin{equation}
P_{\mathrm{DC,TX}} = P_{\mathrm{DC,VCO}} + P_{\mathrm{DC,mixer}} + P_{\mathrm{DC,PA}},
\end{equation}
\begin{equation}
P_{\mathrm{DC,RX}} = P_{\mathrm{DC,LNA}} + P_{\mathrm{DC,ED}}.
\end{equation}
 
On the transmitter side, since the PA is connected to the on-chip antenna and we consider full co-integration of both elements, the transmitted RF power is assumed to be equal to the PA output power, i.e., $P_{\mathrm{t}} = P_{\mathrm{out,PA}}$. On the receiver side, the required input signal level is dictated by the sensitivity expression in Equation~\eqref{eq:sensitivity_rewritten}, which maps the target BER and data rate to a minimum SNR requirement and thereby sets the operating conditions of the LNA and the ED. 

To identify representative system-level trends, the optimal data points from the surveyed literature are used for each RF sub-block. For a given operating frequency, the design metrics most directly linked to power consumption are first identified. As an illustrative example, for the PA, the Power-Added Efficiency (PAE) is chosen as the primary open parameter. Since a higher PAE corresponds to a lower DC power consumption for a given output power, the data points exhibiting the highest PAE at each frequency are interpreted as best-in-class realizations among the reported designs. Frequency-dependent behavioral models are then derived by fitting trend lines to these upper-envelope data points. This approach captures optimistic yet realistic performance limits while reducing the impact of non-optimized implementations.
\vspace{-0.1cm}
\section{Transmitter Power Consumption Modeling}\label{Section: Methodology_TX}
Among the RF front-end components of the transmitter, PAs are typically the most energy-demanding blocks. However, accurately modeling PA power consumption remains challenging, as it strongly depends on factors such as biasing class, semiconductor technology, and operating conditions~\cite{egan2004practical}. To address this complexity in the context of WNoC systems, we develop a frequency-dependent behavioral model for estimating the transmitter’s DC power consumption ($P_{\mathrm{DC,TX}}$), starting with the PA as the primary contributor. Then, we also add LO and mixer considerations.

\vspace{-0.2cm}

\subsection{Power Amplifier}
To model the PA behavior for WNoC systems, the proposed method depends on data obtained from the PA designs presented in~\cite{wang2020power}. This survey aggregates circuits published from 2000 to the present, covering a broad spectrum of carrier frequencies and commercial semiconductor technologies, such as CMOS, SiGe, LDMOS, GaN, InP, and GaAs. In this work, CMOS technology is selected because it allows for monolithic integration of the TRX within computing systems such as CMOS-based SoC platforms.

The PA power consumption $P_{\mathrm{DC,PA}}$ is calculated by using the Power Added Efficiency (PAE) as
\begin{equation}
P_{\mathrm{DC,PA}}^{\mathrm{mW}}(f) = \frac{P_{\mathrm{out}}^{\mathrm{mW}} - P_{\mathrm{in}}^{\mathrm{mW}}}{\mathrm{PAE}(f)},
\label{eqn:Pdc}
\end{equation}
where \( f \) denotes the operating frequency, and \( P_{\mathrm{out}} \) and \( P_{\mathrm{in}} \) represent the RF output and input power, respectively. In particular, we obtain a fitting of $PAE(f)$ from the PA survey and leave \( P_{\mathrm{out}} \) and \( P_{\mathrm{in}} \) as open parameters. The rationale for adopting PAE as the basis for this model is the extensive availability of PAE data for CMOS-based amplifiers operating across diverse frequency bands, enabling reliable frequency-dependent power estimation.

\begin{figure}[!b]
\vspace{-0.6cm}
\centering
\includegraphics[width=83mm]{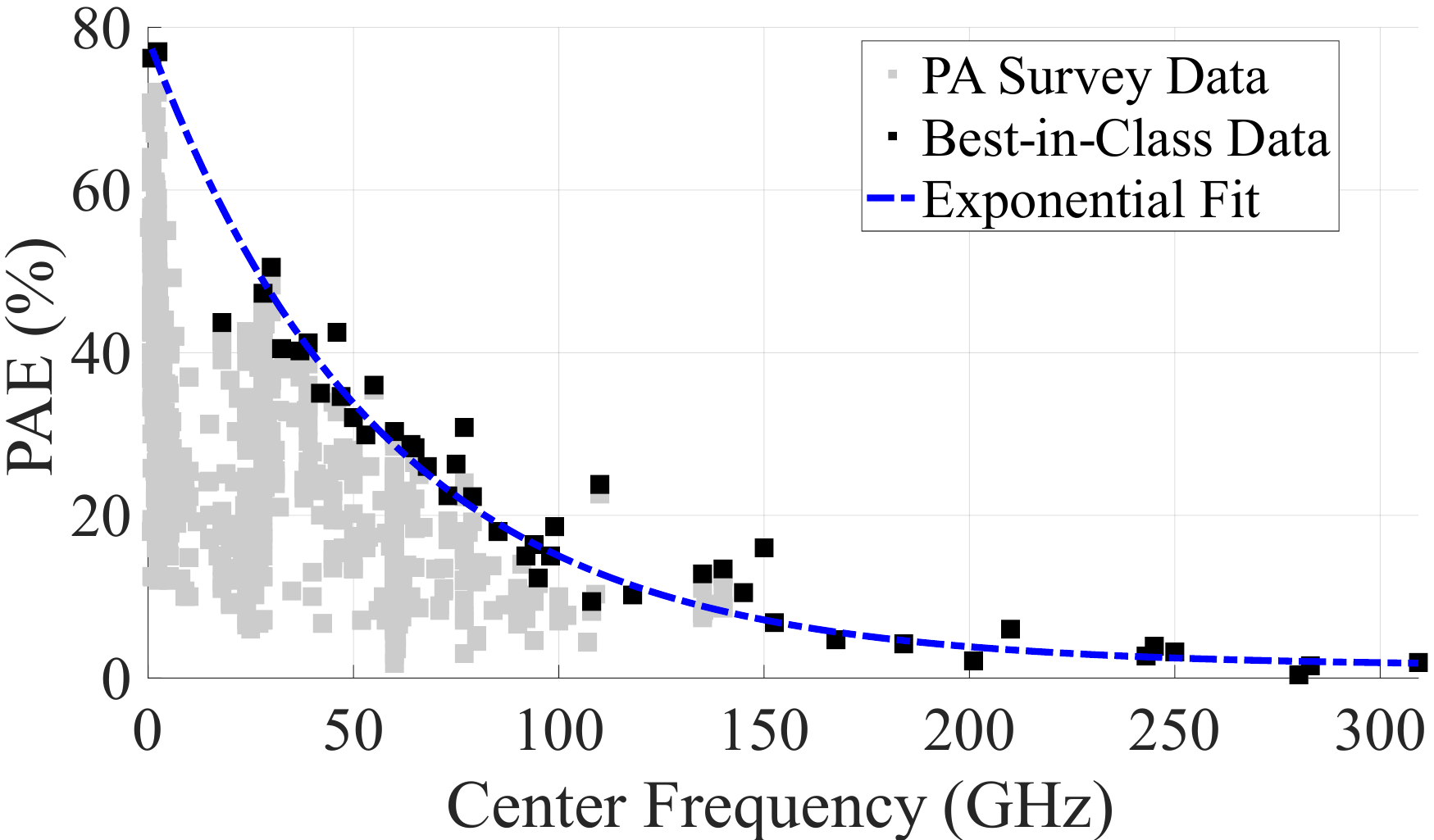} 
\includegraphics[width=83mm]{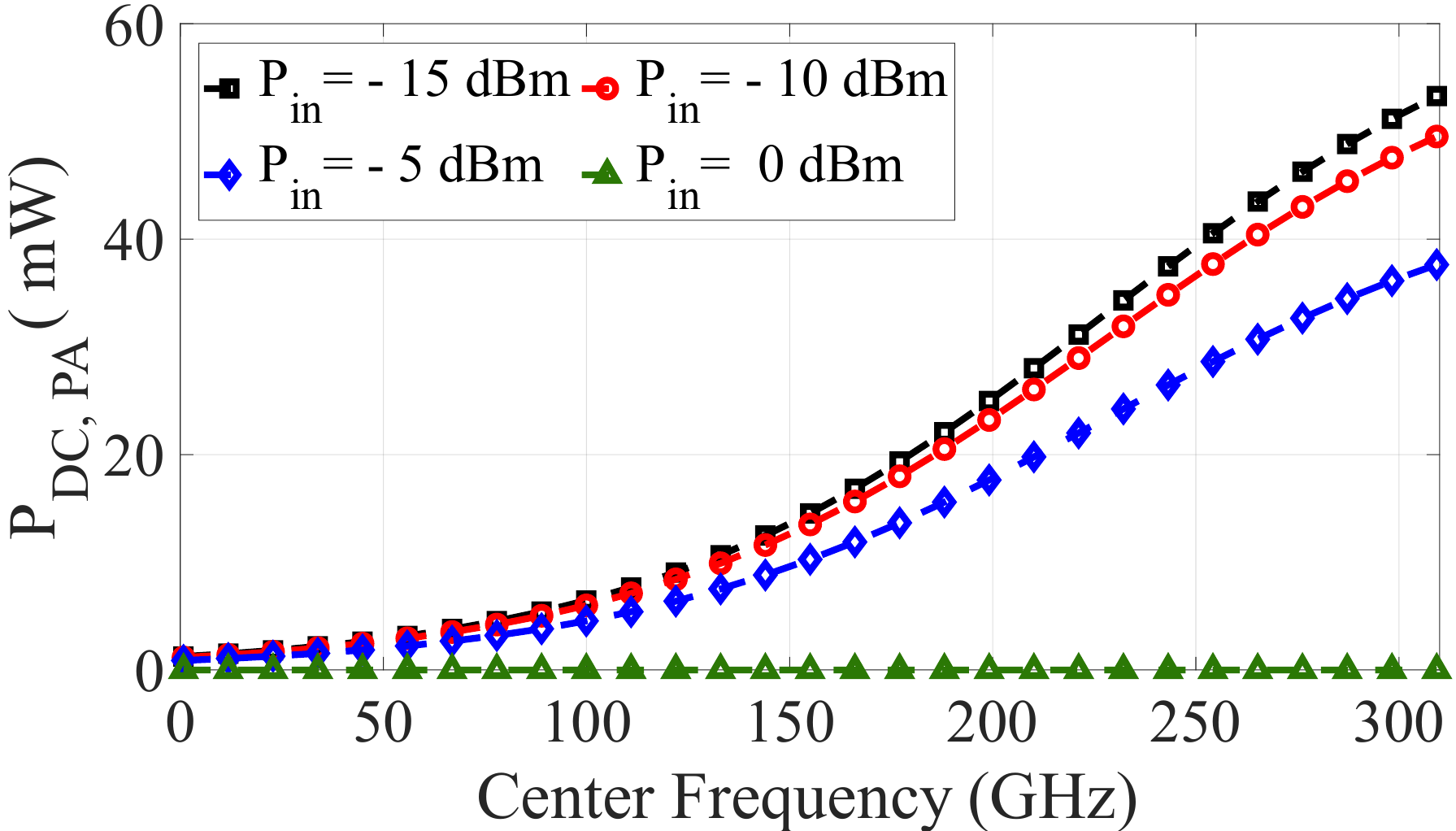} 
\vspace{-0.2cm}
\caption{Top: frequency-dependent PAE model derived from best-in-class data in~\cite{wang2020power}. 
Bottom: resulting \(\mathrm{P_{DC, PA}}\) for multiple \(\mathrm{P_{in}}\) configurations.}
\label{fig:PAE_dc_Power_PA_Survey}
\vspace{-0.2cm}
\end{figure}

The relationship between PAE and operating frequency is extracted from best-in-class data points reported in the survey (highlighted as black squares in Fig.~\ref{fig:PAE_dc_Power_PA_Survey}, top). Best-performing data are selected across frequencies to capture the upper efficiency envelope of state-of-the-art PA implementations, minimizing the influence of architecture-specific limitations and early-stage prototypes. An exponential curve-fitting procedure was then applied to this data, yielding a frequency-dependent model of PA power usage with a coefficient of determination of $R^{2} = 0.95$ over the range $1 \leq f \leq 310~\mathrm{GHz}$. The reduction of efficiency as frequency increases can be explained by higher losses in passive components and the progressive approach to $f_{T}$ and $f_{max}$ of CMOS, which leads to reduced transistor gain.

To test the model, multiple values of the input and output power of the PA are considered. To choose them, we note that in the specific case of on-chip wireless interconnects, where communication distances are inherently short, high-power PAs are unnecessary. Accordingly, \(P_{\mathrm{out,PA}}\) is fixed at 0 dBm, ensuring the model reflects the use of low-power amplifiers suited to WNoC applications. Also, \(P_{\mathrm{in}}\) is varied from $-$15 to 0 dBm. This range encompasses both low-power PA use cases reported in the PAs survey and scenarios that require no amplification, i.e., 0 dB gain. 
The results obtained with Eqn.~\eqref{eqn:Pdc} and illustrated in the bottom plot of Fig.~\ref{fig:PAE_dc_Power_PA_Survey}, show how the PA power consumption increases sharply with the operating frequency. At $f = 310~\mathrm{GHz}$ and $\mathrm{P_{in}} = -15~\mathrm{dBm}$, a 15 dB gain corresponds to a power consumption of approximately 53 mW, which is significant for short-range applications.

\subsection{Local Oscillator}\label{subsection: OSc_Behavioral_Modeling}

The oscillators are another key sub-block in RF TRXs and can account for a non-negligible portion of power consumption. As reported in~\cite{yu2014architecture}, the contribution of this sub-block becomes particularly significant under BPSK and QPSK modulation schemes, where their power demand can even surpass that of the PA due to the need for a power-hungry Phase-Locked Loop (PLL). In simpler architectures, such as OOK, a Voltage Controlled Oscillator (VCO) is typically employed only on the transmitter side to generate the carrier frequency for up-conversion, without requiring an additional PLL for frequency synthesis. Despite VCO's modest power, it still contributes significantly to the total TRX energy.

To model the power consumption of VCOs, this study focuses on fundamental oscillator designs due to their higher energy efficiency~\cite{underhill1992fundamentals}, an important consideration in energy-constrained applications such as WNoC. 
The modeling methodology follows the same principles outlined for PAs. The initial parameter extraction is based on a survey of high-frequency oscillator designs~\cite{wang2020power}. To extend the frequency coverage and enhance the representativeness of the dataset, additional data from studies on fundamental oscillators operating up to 272 GHz were included, complemented by our own survey-based dataset~\cite{shahmoradi2025vcosurvey}. In total, the proposed survey-based modeling relies on 129 data points. This approach ensures both wide frequency coverage and robust trend analysis.

The power consumption of the oscillator $P_{\text{DC,VCO}}$ is quantified using the DC-to-RF efficiency metric, which is defined as the ratio of RF output power $P_{\text{RF,VCO}}$ to the corresponding DC power consumption at a given frequency $P_{\text{DC,VCO}}$. To obtain the power consumption across frequencies, we use the following expression derived from the efficiency metric, 
\begin{equation}
P_{\text{DC,VCO}}(f) = \frac{P_{\text{RF,VCO}}}{\text{DC-to-RF Efficiency}(f)}.
\label{eqn:Osc_dcRF}
\end{equation}

\begin{figure}[!t]
\centering
\includegraphics[width=83mm]{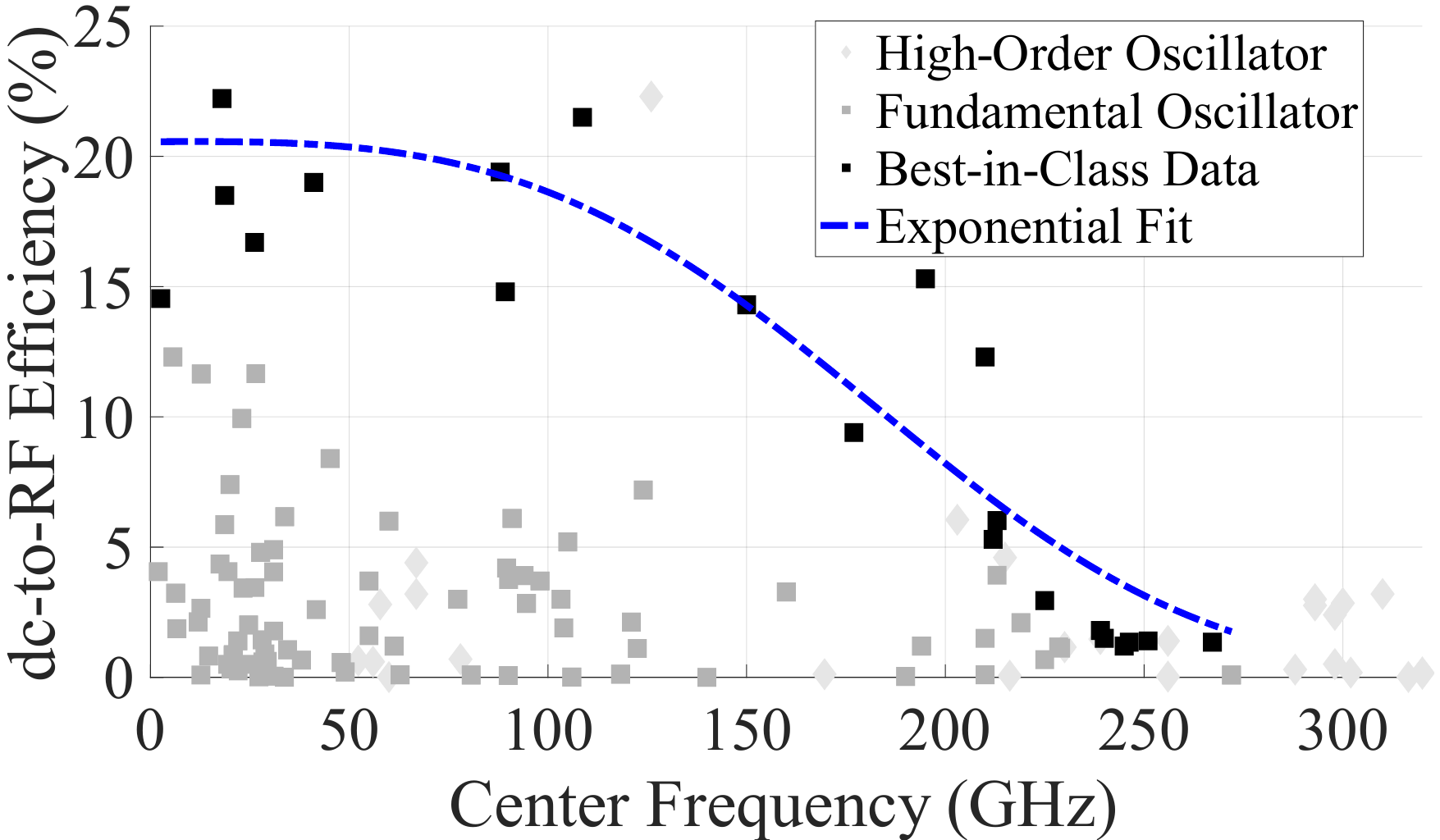} 
\includegraphics[width=83mm]{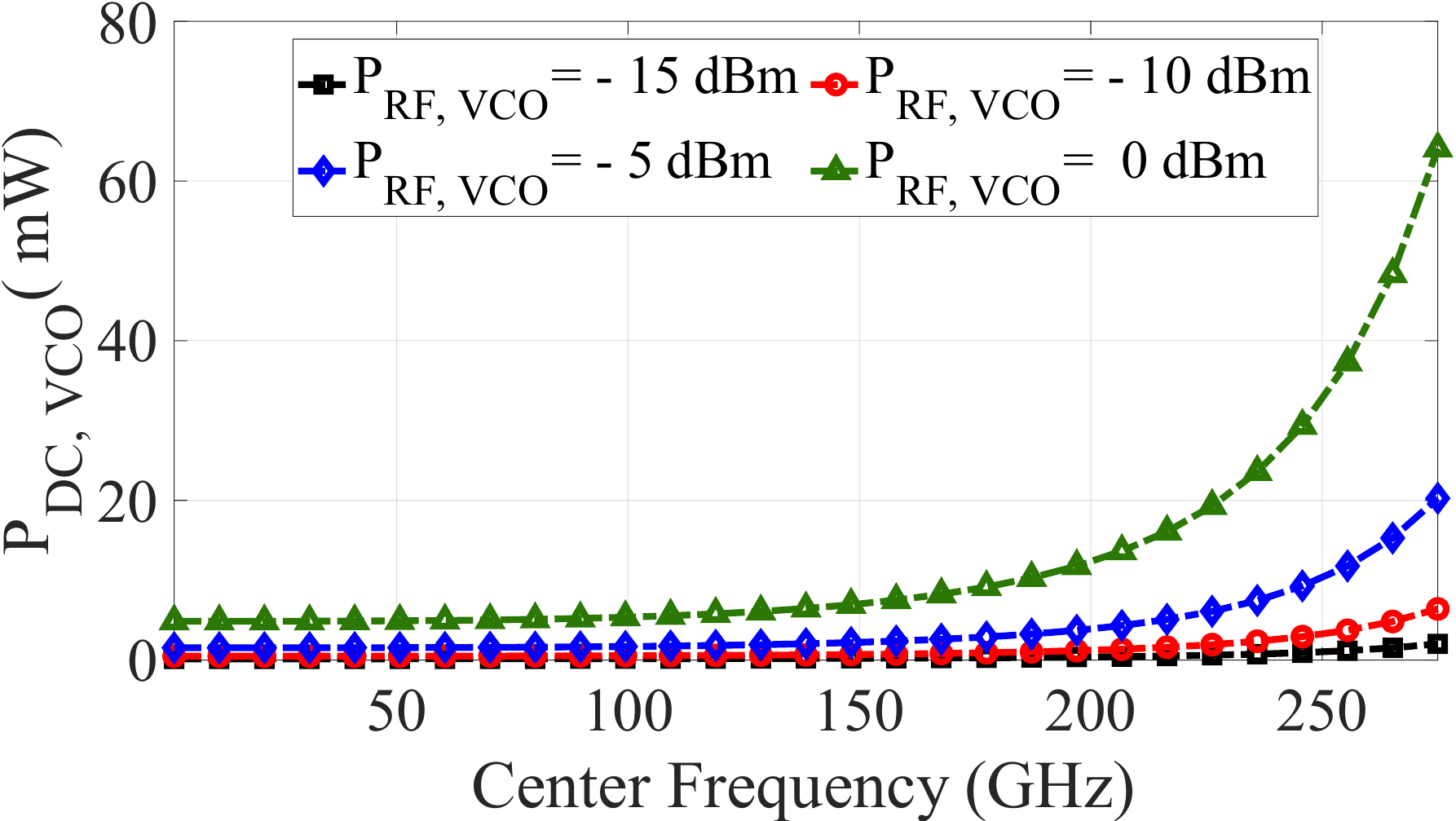} 
\caption{Frequency-dependent oscillator characterization. 
Top: DC-to-RF efficiency model versus operating frequency. 
Bottom: corresponding oscillator DC power \(\mathrm{P_{DC,VCO}}\) for different output-power levels \(\mathrm{P_{RF,VCO}}\).}
\label{fig:dc_Power_Oscillator_Survey}
\vspace{-0.3cm}
\end{figure}

In this case, the survey data is used to obtain a formula for $\text{DC-to-RF Efficiency}(f)$ through regression and leaving the RF output power value of ($P_{\text{RF,VCO}}$) as an open parameter. The regression analysis over best-in-class data across the frequency range \(2 \leq f \leq 272~\mathrm{GHz}\), illustrated in the top plot of Fig.~\ref{fig:dc_Power_Oscillator_Survey}, yields an exponential power trend for fundamental oscillators with an \(R^2\) value of 0.83. The reduction of efficiency can be explained by the degradation of the resonator quality factor ($Q$) at higher frequencies.

Using the DC-to-RF efficiency definition from Eqn.~\eqref{eqn:Osc_dcRF}, the behavior of \(\mathrm{P_{DC,VCO}}\) is examined. As shown in the bottom plot of Fig.~\ref{fig:dc_Power_Oscillator_Survey}, fundamental oscillators in WNoC systems generally consume less than 10 mW of DC power for frequencies below 200 GHz, making them suitable for low-power applications. However, power consumption can rise substantially at sub-terahertz frequencies, negatively impacting efficiency and increasing system cost. 

\vspace{-0.2cm}
\subsection{OOK Modulator}
In this work, OOK modulation is implemented using a mixer-based architecture, where the mixer directly performs the modulation. Gilbert-cell mixers are the most widely used topology for this purpose and can be classified as either passive or active~\cite{razavi2021fundamentals}. In WNoC systems, passive mixers are preferred for power modeling due to their lower power consumption and inherent compatibility with OOK operation~\cite{thomas2024400, kim20172, im2019220, panes2015energy}.
Active mixers, on the other hand, can be interpreted as a passive switching stage followed by amplification, whose power contribution can be effectively captured within the PA model. This abstraction avoids double-counting and maintains consistency to minimize overall TRX power consumption.

Our model takes data from a recent surveys on mixer performance~\cite{zhang2022optimal, shahmoradi2025mixersurvey}, which together offer 180 data points of power consumption across a wide range of frequencies. Using these benchmarks, the mixer's power consumption as a function of the operating frequency can be modeled as follows.

For mixers, a usual performance metric is the conversion gain (CG) or its counterpart, conversion loss (CL), which is typically formulated as
\begin{equation}
\text{CG (dB)} = 10 \cdot \log_{10} \left( \frac{P_{\text{RF,out}}}{P_{\text{BB,in}}} \right) = - \text{CL (dB)},
\label{eqn:Mix_CG}
\end{equation}
where $P_{\text{RF,out}}$ is the required RF power at the output of the mixer and $P_{\text{BB,in}}$ is the baseband power at its input. 

\begin{figure}[!t]
\centering
\includegraphics[width=83mm]{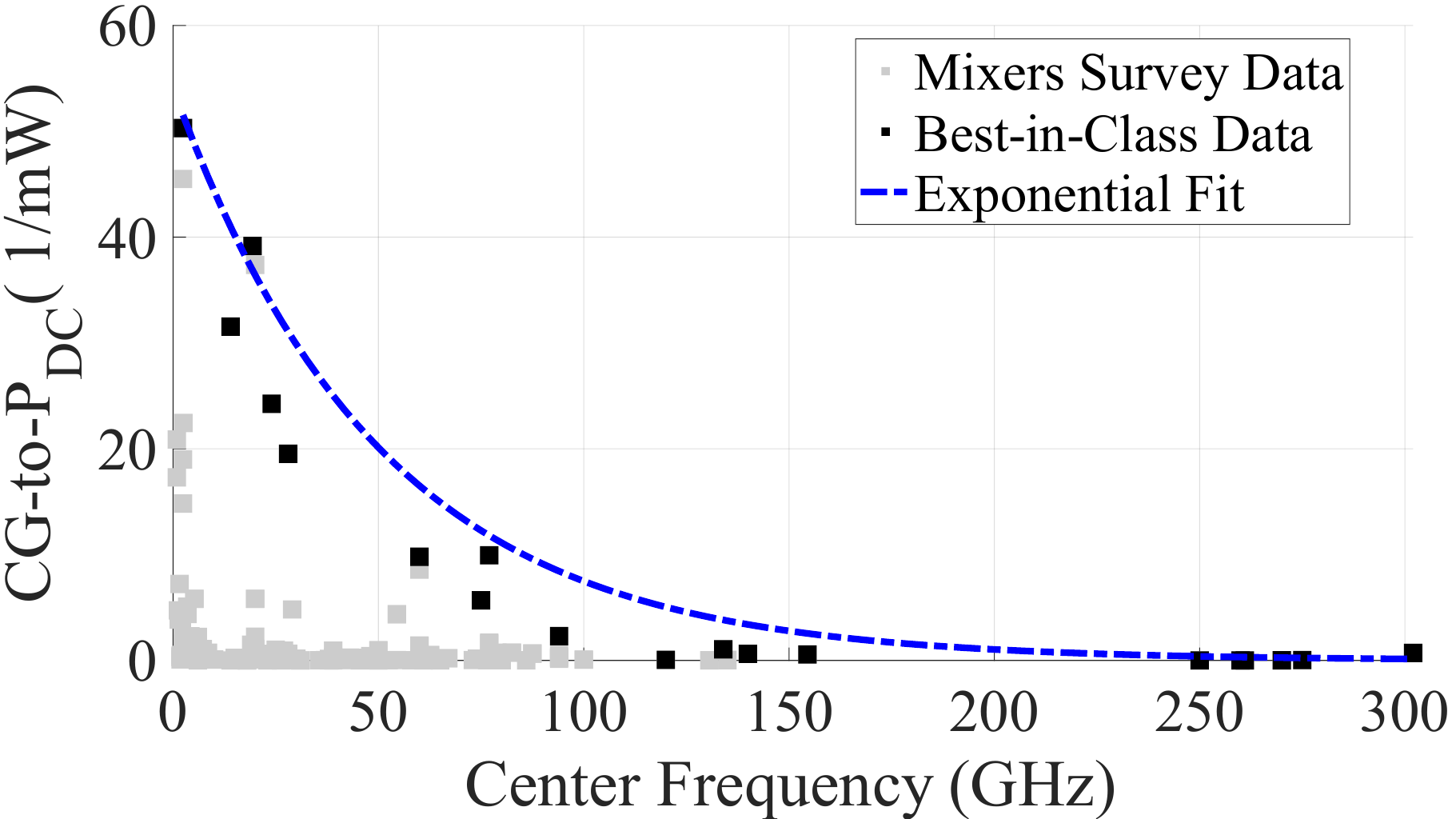} 
\includegraphics[width=83mm]{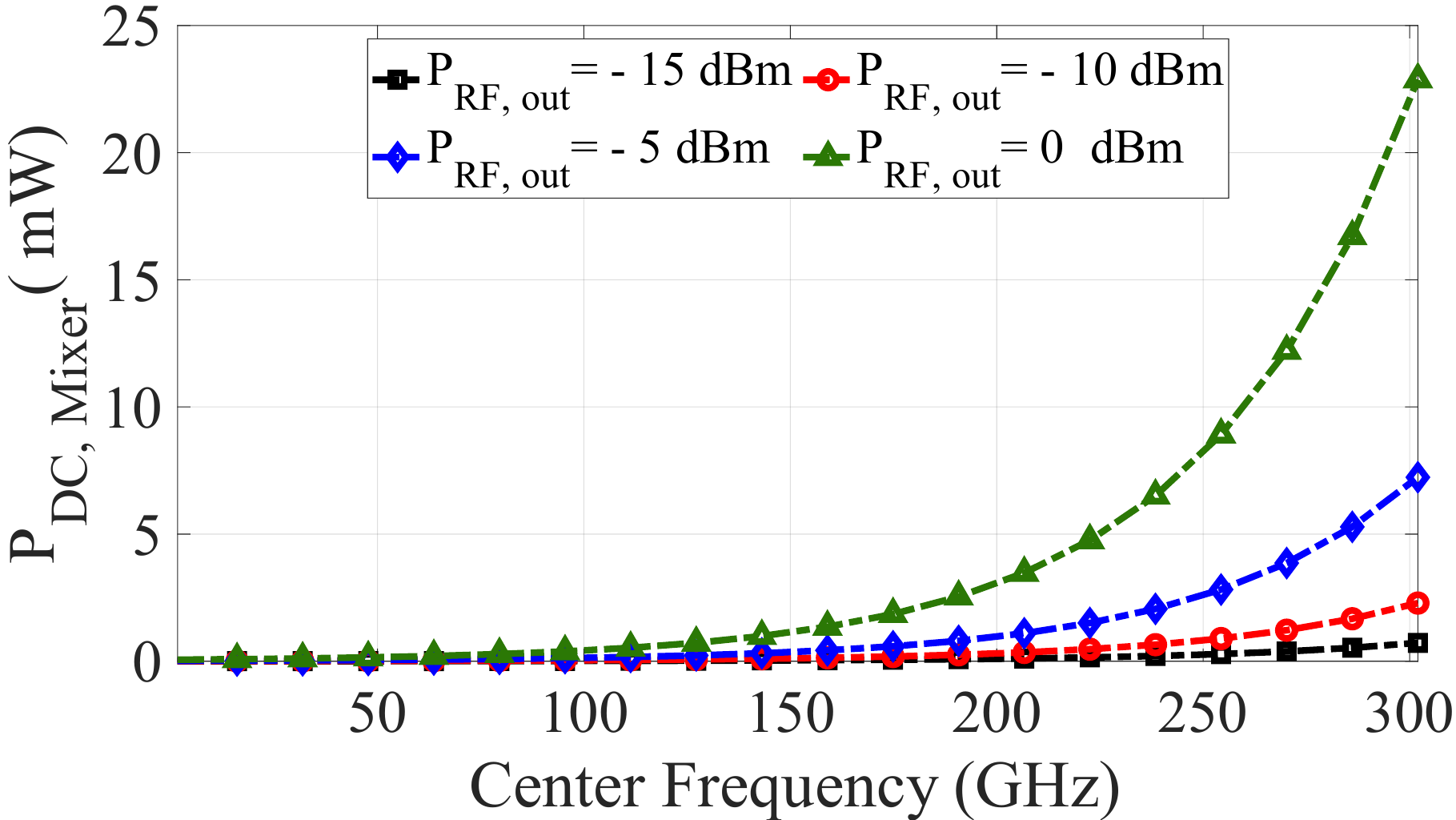} 
\vspace{-0.2cm}
\caption{Top: mixer DC power consumption versus operating frequency, obtained from the aggregated survey data. Bottom: corresponding fitted model for \(\mathrm{P_{DC, Mixer}}\).}
\vspace{-0.2cm}
\label{fig:Mixer_CG_Over_Pdc}
\end{figure}

The CG is typically correlated with the power consumption of the mixer $P_{DC,mixer}$ because, fundamentally, CG is proportional to \(R_L \cdot \mathrm{P_{DC}} / V_{DD}\) as analyzed in~\cite{razavi2021fundamentals}. Therefore, several works consider the ratio $\text{CG-to-P}_{DC} = \tfrac{10^{CG/10}}{P_{DC,mixer}}$ as a valid figure of merit. Evaluating this expression across frequencies, we can obtain a model of the mixer's power consumption as
\begin{equation}
P_{DC,mixer}(f) = \frac{10^{CG/10}}{\text{CG-to-P}_{DC}(f)}.
\label{eqn:P_DC_Mix_CG}
\end{equation}

Our methodology obtains $\text{CG-to-P}_{DC}(f)$ through regression using the survey data and leaves the CG as an open parameter that depends on $P_{\text{RF,out}}$ and $P_{\text{BB,in}}$. On the one hand, the value of \(\mathrm{P_{RF,out}}\) is left as an open parameter that matches the assumed input power of the PA. On the other hand, for the value of $P_{\text{BB,in}}$, it is worth noting that the baseband signal in WNoC systems employing OOK modulation is typically used to control the switching behavior of the mixer, effectively modulating the carrier. Given the low-power nature of on-chip wireless communication, the required $P_{\text{BB,in}}$ is relatively low, typically ranging from $1 ~\upmu\mathrm{W}$ to $1~\mathrm{mW}$ depending on the circuit design and technology node~\cite{gade2019energy, mondal2017energy}. 
In particular, our study assumes a fixed $\mathrm{P_{BB, in}}$ of 0.1 mW.

The regression of the CG-to-\(\mathrm{P_{DC}}\) ratio versus frequency, based on best-in-class data, is presented in the top panel of Fig.~\ref{fig:Mixer_CG_Over_Pdc}. An exponential function is fitted over the interval \(0.9 \leq f \leq 302 \,\mathrm{GHz}\), achieving an \(R^2\) value of 0.97. The corresponding evolution of \(\mathrm{P_{DC,mixer}}\) is depicted in the bottom panel of Fig.~\ref{fig:Mixer_CG_Over_Pdc}. These findings indicate that the mixer power consumption remains relatively modest at lower frequencies but can become substantial at high power levels and as the frequency approaches \(f = 300 \,\mathrm{GHz}\). Such a reduction of conversion efficiency can be explained by limitations in switching speed and reduced efficiency of the LO drive as frequency increases.


\section{Receiver Power Consumption Modeling} \label{Section: Methodology_RX}
In wireless communication networks, it is widely acknowledged that overall energy consumption is largely dominated by the transmission circuitry. However, in WNoC systems, the energy requirements of the receiver-side may gain particular relevance. This change in emphasis is driven by the presence of a large number of concurrently active receiver units, which could render the power consumption of a single broadcast message excessive. To reduce the receiver power consumption, a non-coherent receiver architecture for OOK is considered, composed of an LNA plus a demodulator implemented by an analog ED. Next sections describe the models for the LNA and ED, respectively. 
\vspace{-0.2cm}

\subsection{Low Noise Amplifier}\label{subsection: LNAs_Methodology_Modeling}

First and foremost, to model the power consumption of LNAs in a principled and technology-independent manner, we follow the same systematic methodology adopted for the transmitter sub-blocks. Our analysis builds upon the extensive survey in~\cite{belostotski2020low}, which compiles more than 600 CMOS LNA implementations and extracts over 35 figures of merit. From this dataset, the metrics most relevant to WNoC receiver design, namely DC power consumption, gain, and noise factor (F), serve as the foundation of our modeling approach.

\begin{figure}[!t]
\centering
\includegraphics[width=83mm]{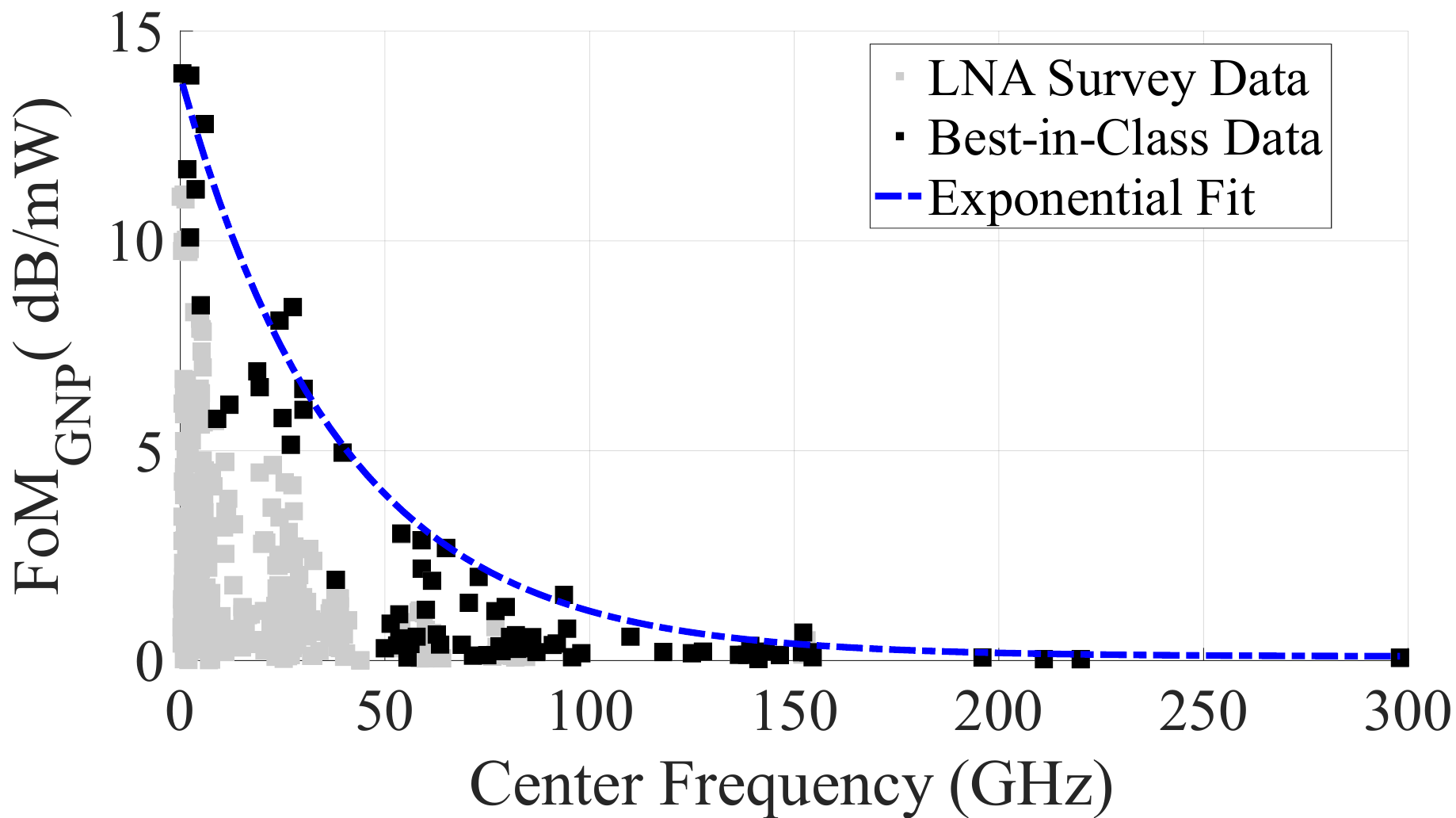} 
\includegraphics[width=83mm]{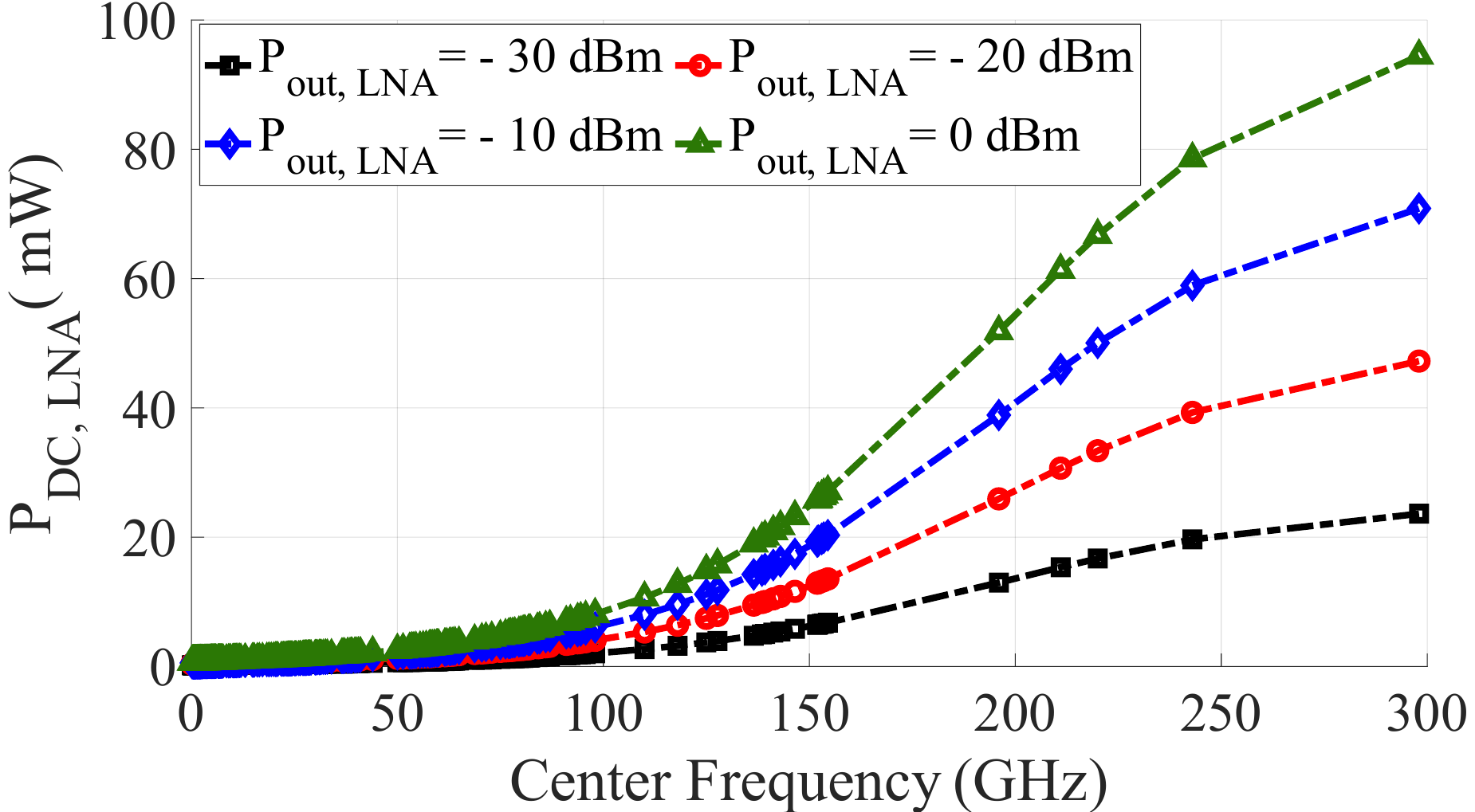} 
\vspace{-0.2cm}
\caption{LNA power modeling in the WNoC receiver chain. 
Top: Exponential regression of \(\mathrm{FOM}_{\mathrm{GNP}}\) based on best-in-class data. 
Bottom: Modeled power consumption for gains of 10–40 dB at NF = 7 dB and $P_\mathrm{in}=-40$ dBm.}\label{fig:LNA_FOM}
\vspace{-0.2cm}
\end{figure}

To facilitate a fair comparison across different designs, we adopt the figure of merit introduced in~\cite{belostotski2020low}, 
\begin{equation}
    \mathrm{FOM_{GNP}}
    = \frac{\mathrm{Gain [dB]}}{(F-1) P_\mathrm{DC,LNA} [\mathrm{mW}]},
\end{equation}
where $F$ represents the LNA noise factor in linear units. To evaluate how the LNA power consumption varies with frequency, a regression is applied to the FOM, and $P_\mathrm{DC,LNA}$ is then obtained by solving the resulting expression as
\begin{equation}
    P_\mathrm{DC,LNA}(f) = \frac{\mathrm{Gain [dB]}}{(F-1) \mathrm{FOM_{GNP}}(f)}.
\end{equation}

To derive the behavioral model of the LNA, an exponential regression was employed to correlate \( \text{FOM}_{\text{GNP}} \) with operating frequency. The resulting trend is depicted in the top plot of Fig.~\ref{fig:LNA_FOM}. A negative exponential fitting is obtained with the best-in-class data, achieving an \(R^2\) of 0.86. Such a decrease in efficiency is due to reduced transconductance efficiency and higher parasitic capacitances, leading, in general, to lower gain and higher noise figure at high frequencies.

To illustrate our model, consider the case where the received power is $-$40 dBm and the required $P_\mathrm{out,LNA}$ is swept from -30 dBm to 0 dBm, corresponding to LNA gains ranging from 10 dB to an extreme case of 40 dB. The bottom plot of Fig.~\ref{fig:LNA_FOM} shows the resulting power consumption assuming a noise figure held constant at 7 dB. At the highest considered gain of the estimated power consumption at the extreme case of 40 dB gain goes from 1.45 mW at 28 GHz to 78.5 mW at 245 GHz. By contrast, when the NF is improved to 2 dB while maintaining the same gain (not shown in the figure), the power consumption increases significantly to 9.95 mW and 535 mW at the same frequencies. This demonstrates a fundamental trade-off: achieving a lower noise figure and higher gain simultaneously leads to much higher power consumption.

\subsection{OOK Demodulator}
Within non-coherent OOK TRXs in the WNoC context, ED power modeling completes the receiver model and, consequently, the overall TRX power analysis. Prior work has demonstrated that ED-based architectures enable ultra-low-power operation. For instance, a fully integrated OOK receiver consuming only \(178~\upmu\mathrm{W}\) is reported in~\cite{boora2024ultra}, while a 60~GHz OOK demodulator based on envelope detection achieves multi-Gb/s data rates with low RF complexity~\cite{yu201518}. Moreover, the eWake architecture in~\cite{kazdaridis2021ewake} shows that semi-passive wake-up receivers can maintain adequate sensitivity at extremely low power levels.

\begin{figure}[t]
\centering
\includegraphics[width=83mm]{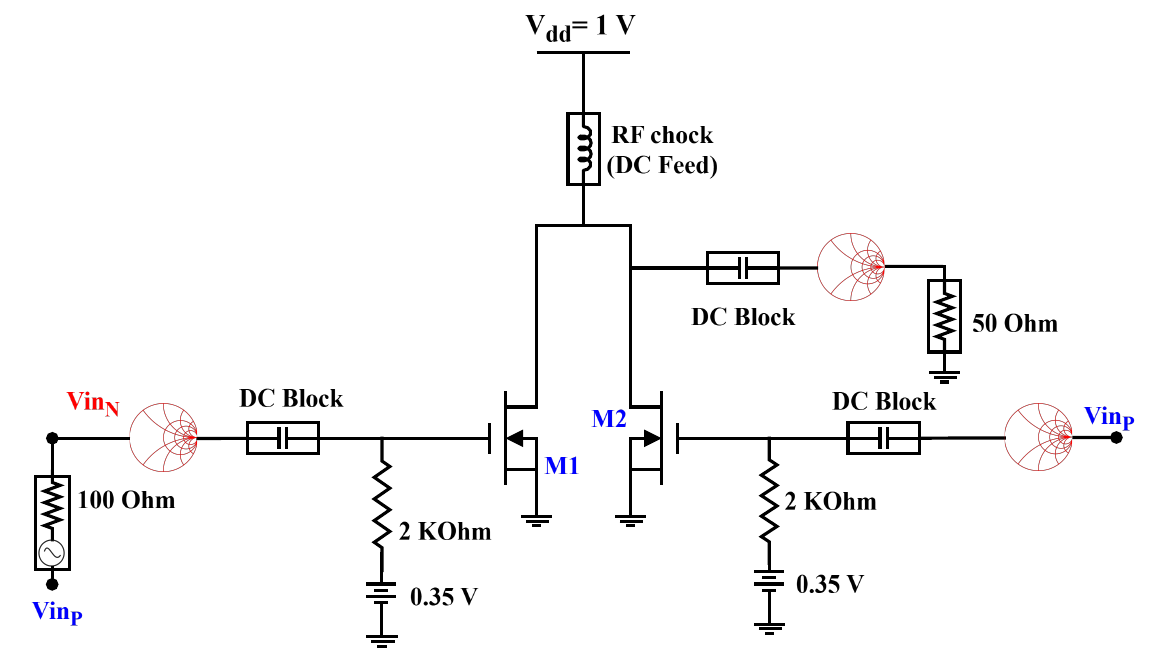} 
\vspace{-0.2cm}
\caption{Schematic of the ED implemented in TSMC 65 nm CMOS technology.}\label{fig:Envelope_Detector_Schematic}
\end{figure}

Although the ED generally exhibits low power consumption, under certain operating conditions and frequencies its power can become comparable to that of the LNA. Due to the limited availability of comprehensive survey data, its characterization in this work relies on simulation. Fig.~\ref{fig:Envelope_Detector_Schematic} shows a conventional ED architecture used in multiple works on high-speed transceivers for low-power short-range communications \cite{peng202426, nakajima201423gbps}. Such ED architecture is implemented Keysight ADS and simulated assuming a Taiwan Semiconductor Manufacturing Company (TSMC) 65 nm CMOS technology, and using 60 nm RF NMOS transistors within the receiver chain. For each discrete frequency point, separate matching networks have been design to ensure that matching does not impact power consumption as frequency is increased.

\begin{figure}[t]
\centering 
\includegraphics[width=83mm]{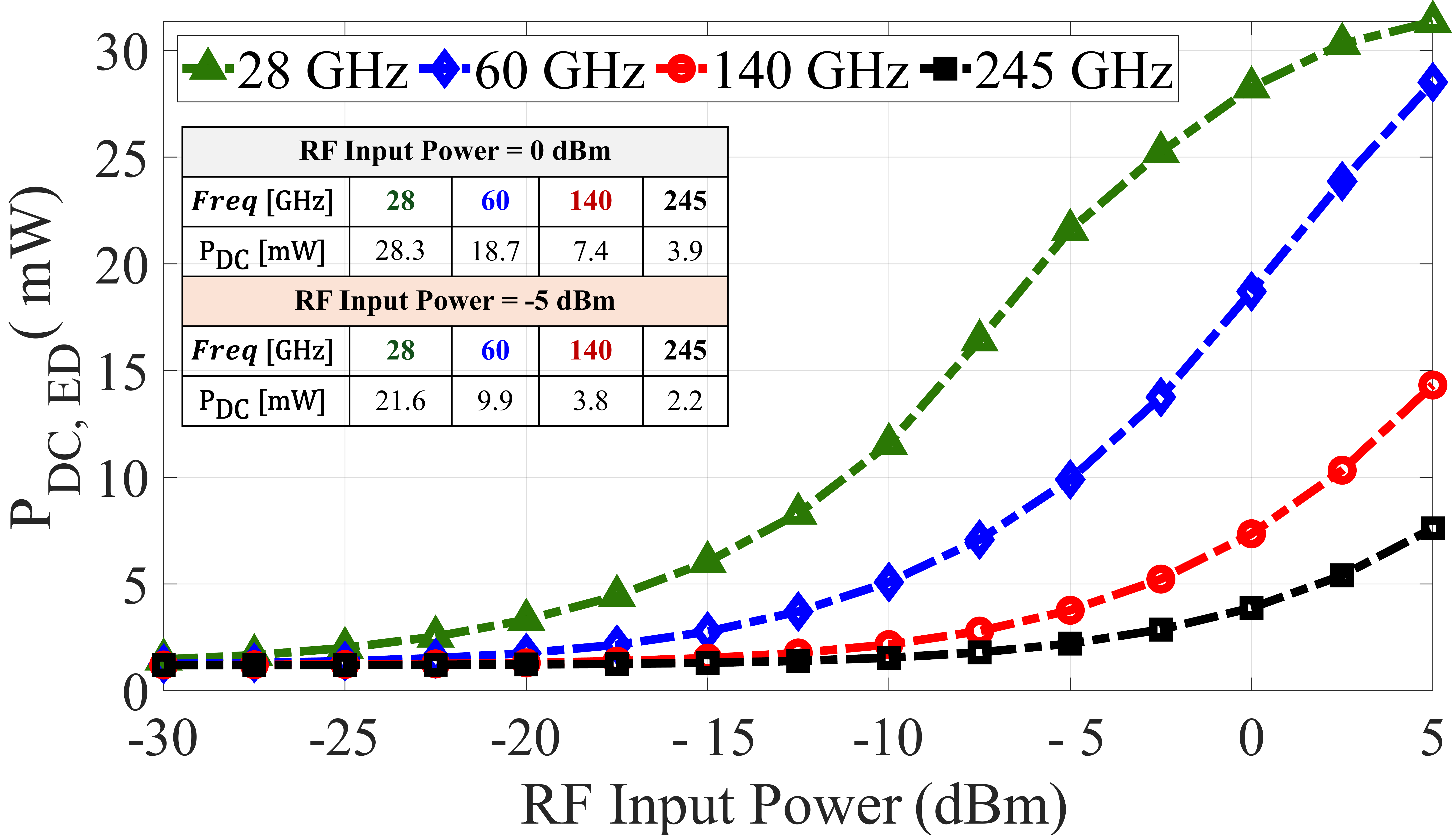} 
\includegraphics[width=83mm]{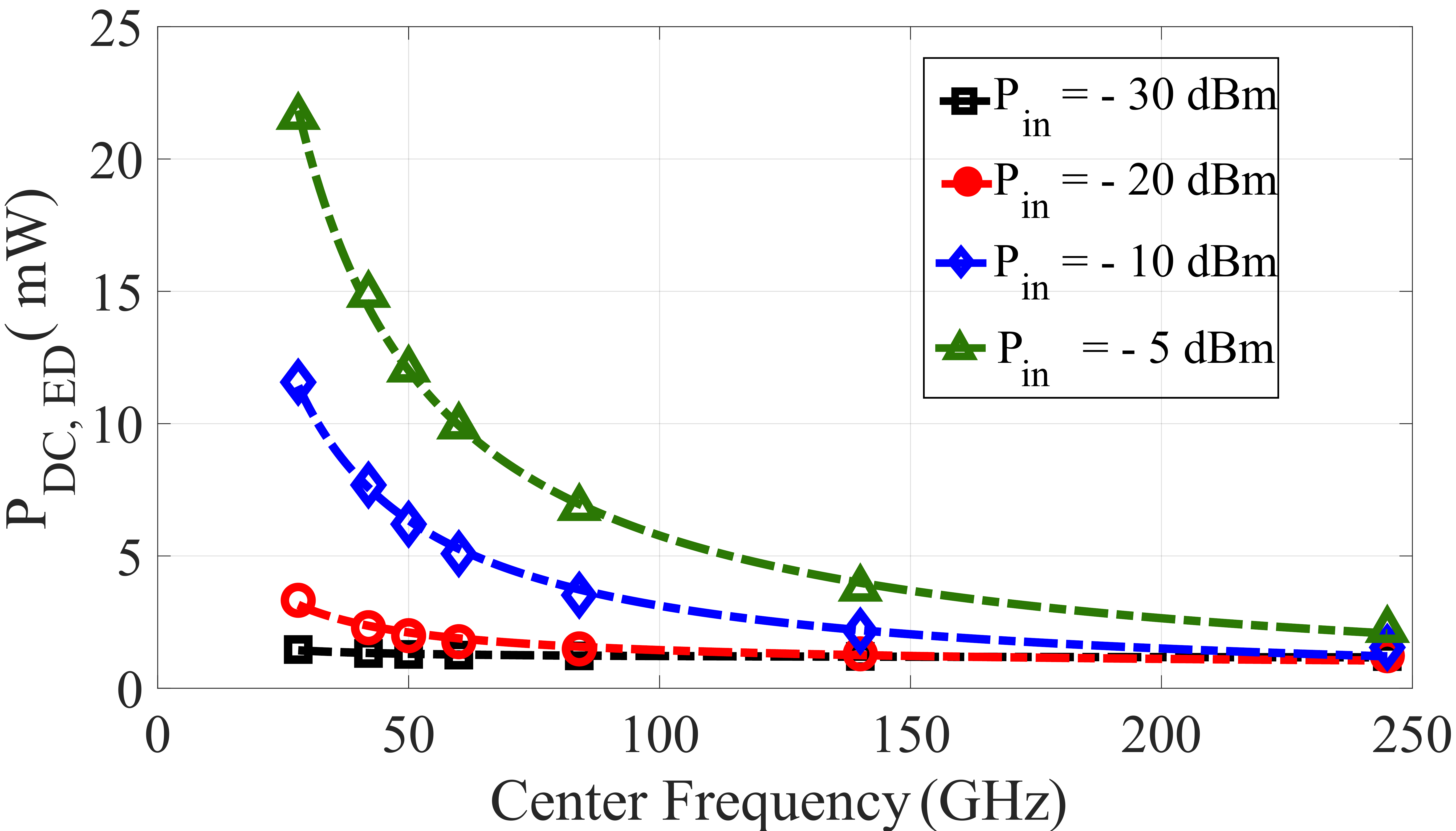} 
\vspace{-0.2cm}
\caption{Behavior of the derived model of \(\mathrm{P_{DC, ED}}\). Top: variation with RF input power at a fixed frequency. Bottom: variation with frequency at a fixed RF input power.}
\vspace{-0.2cm}
\label{fig:Envelope_Detector_Model}
\end{figure}

Fig.~\ref{fig:Envelope_Detector_Model} presents the corresponding first-order power consumption estimates obtained in the simulation. As shown in the top plot, assuming an RF input power of $-$5 dBm driven by the LNA output, the simulated DC power of the ED is approximately \(21.6\), \(9.9\), \(3.8\), and \(2.2\,\mathrm{mW}\) at \(28\), \(60\), \(140\), and \(245\,\mathrm{GHz}\), respectively. These results define the frequency-dependent operating assumptions of the ED in the proposed model. By further examining Fig.~\ref{fig:Envelope_Detector_Model}, bottom plots, we observe a \(1/f\) trend in the ED power consumption that can be justified theoretically as follows.

Envelope detection relies on maximizing the transistor transconductance \(g_{m2}\), which occurs when the device operates near its threshold voltage \(V_{th}\). Under this biasing condition, the DC supply current, and thus the total power consumption, becomes strongly dependent on the amplitude of the large-signal RF input. When \(V_{\text{bias}} \approx V_{th}\), the negative half-cycle of the sinusoidal input does not significantly contribute to the supply current, and the average DC current can be approximated as
\begin{equation}
I_{\text{DC,avg}} = I_{\text{Q}} + \frac{1}{2\pi} \int_{0}^{\pi} g_{\text{m}} V_{\text{RF,in}} \sin{\theta}   d\theta
\approx \frac{g_{\text{m}} V_{\text{RF,in}}}{\pi} = \frac{I_{\text{RF,out}}}{\pi},
\label{eq:DC_Current}
\end{equation}
where \(I_{\text{Q}}\) is the quiescent current, which is negligible for large input powers. This expression implies a linear relationship between the RF output current and the DC power consumption.

To isolate the frequency dependence, we consider a scenario where a perfect matching condition exists at the input of the ED. Under such circumstances, the RF input current \(I_{\text{RF, in}}\) can be considered nearly frequency-independent. As a result, the frequency characteristics of the power consumption are determined by the ratio \(\frac{I_{\text{RF,out}}}{I_{\text{RF,in}}}\).

Since the ED extracts the baseband component and presents a short-circuit load at the RF carrier, this current ratio scales similarly to the short-circuit current gain \(h_{21}\), where \(f_t\) denotes the transistor transit frequency (the frequency at which \(|h_{21}|=1\)). Given that \(h_{21} \propto \frac{f_t}{f}\), the following proportionality holds,
\begin{equation}
\frac{I_{\text{rf,out}}}{I_{\text{rf,in}}} \propto h_{21} \propto \frac{f_t}{f} \propto P_{\text{DC}}.
\end{equation}
Thus, the inverse dependence of the ED power consumption on frequency arises from the inherent \(1/f\) decay of the transistor current gain under envelope detection biasing conditions, as depicted in Fig.~\ref{fig:Envelope_Detector_Model} (bottom).


\section{Model-Based Energy Efficiency Analysis}\label{Section: Discussion}

Using the power modeling framework we have developed, the analysis begins with a detailed characterization of the individual sub-blocks in the transmitter and receiver RF front-end separately, in Sections~\ref{Section: TX_Modeling} and~\ref{Section: RX_Modeling}, to determine their specific contribution to the total power consumption. Later, by combining all of these individual component models, we obtain a system-level estimate of the overall power consumption of the complete TRX chain in Section~\ref{Section: TRX_Modeling}.

\begin{figure}[!t]
\centering
\includegraphics[width=83mm]{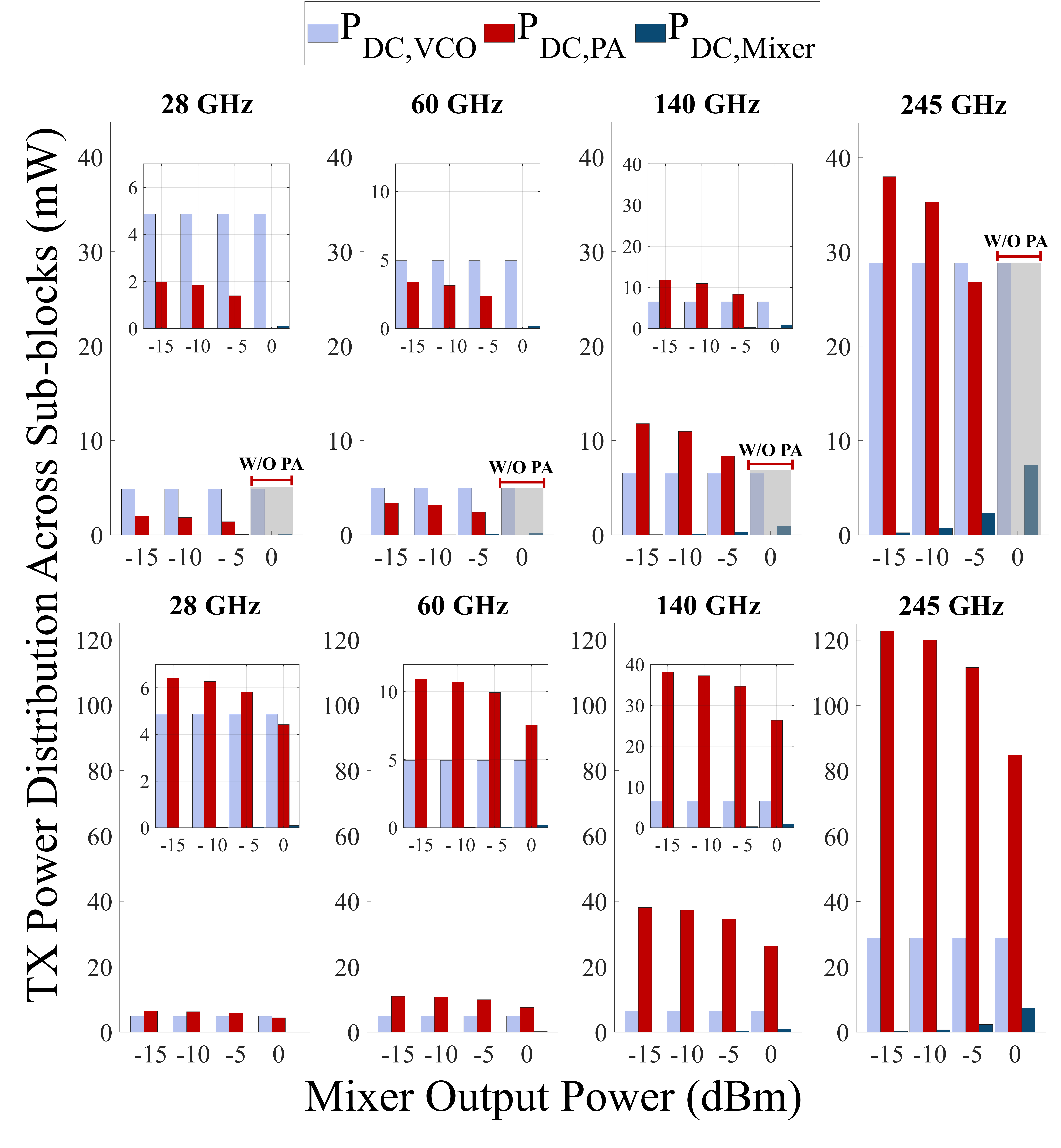}
\vspace{-0.4cm}
\caption{Power consumption of the transmitter front-end for different mixer output power levels with \(\mathrm{P_{IF}} = -5~\mathrm{dBm}\). 
Top: \(\mathrm{P_{out,PA}} = 0~\mathrm{dBm}\) and \(\mathrm{P_{out,VCO}} = 0~\mathrm{dBm}\). 
Bottom: \(\mathrm{P_{out,PA}} = 5~\mathrm{dBm}\) and \(\mathrm{P_{out,VCO}} = 0~\mathrm{dBm}\). 
Gray-shaded regions indicate operating points where no PA is needed.}
\label{fig:TX_Pdc_Model_5dBm_PA_and_0dBm_PA}
\vspace{-0.2cm}
\end{figure}

\subsection{Transmitter Power Consumption}\label{Section: TX_Modeling}
Within the proposed framework, each transmitter sub-block is evaluated as a function of frequency for four mixer output power levels ($P_{\mathrm{RF,out}}=\{-15,-10,-5,0\}$ ~dBm), while the oscillator output power is fixed at 0~dBm. Representative frequencies of 28, 60, 140, and 245~GHz are considered, covering emerging 5G bands, unlicensed mmWave operation, D-band communications, and ISM-band applications, respectively.

\begin{figure}[!t]
\centering
\includegraphics[width=82mm]{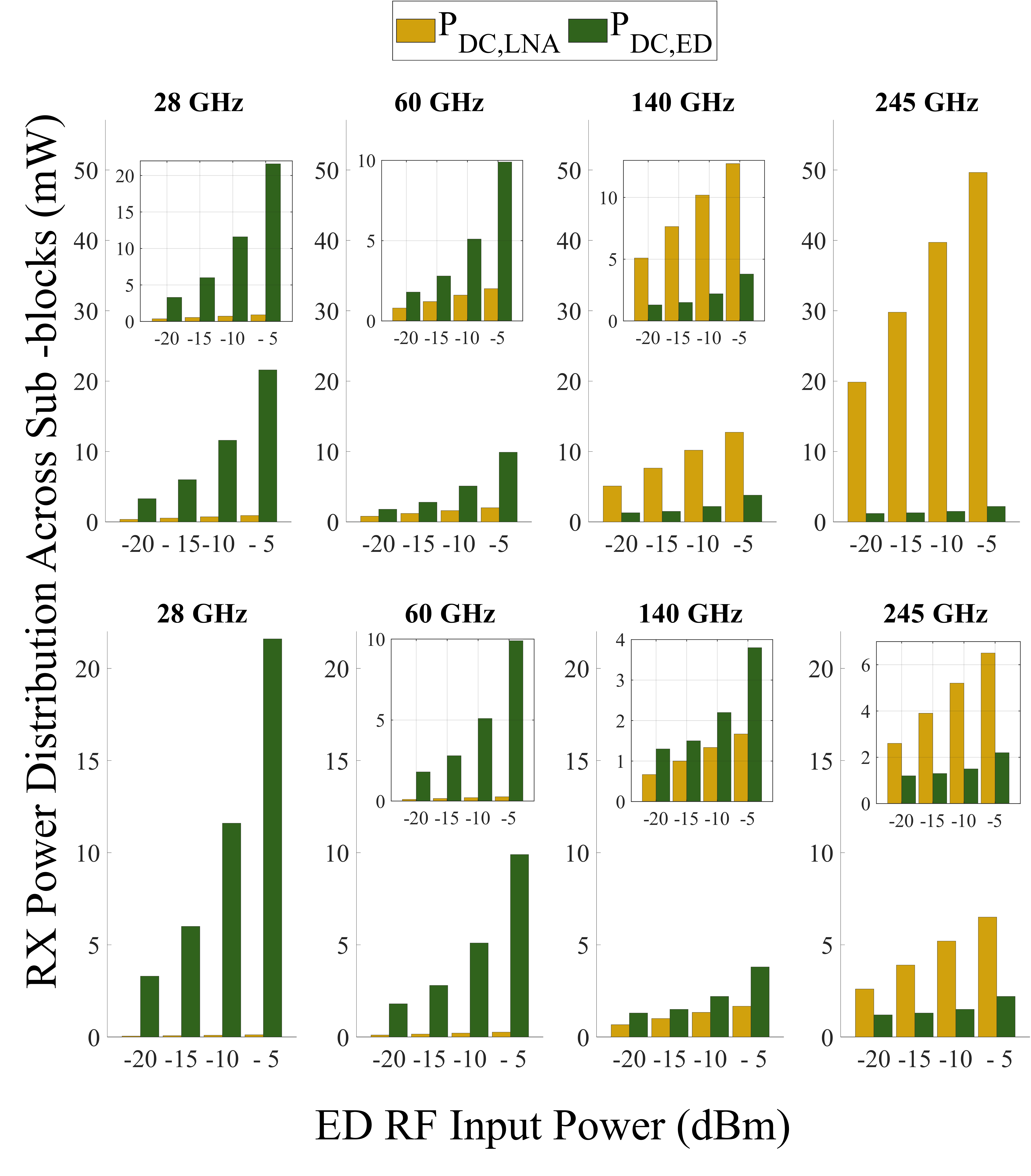}
\vspace{-0.4cm}
\caption{Receiver front-end power consumption with fixed LNA input power of $1 ~\upmu\mathrm{W}$ under varying ED RF input levels. Top: results for $\mathrm{NF}=7$ dB. Bottom: results for $\mathrm{NF}=15$ dB.}
\label{fig:RX_Pdc_Model_5dBm_PA_and_0dBm_PA}
\vspace{-0.2cm}
\end{figure}

We consider two transmitted power levels, 0 $\mathrm{dBm}$ and 5 $\mathrm{dBm}$. With \(\mathrm{P_{out,PA}} = 0~\mathrm{dBm}\), the oscillator dominates power consumption at lower frequencies (Fig.~\ref{fig:TX_Pdc_Model_5dBm_PA_and_0dBm_PA}, top). As frequency increases, both VCO and PA power consumption, but the PA becomes the dominant consumer especially when the mixer output power is low and the PA gain needs to compensate for it. This is more acute for \(\mathrm{P_{out,PA}} = 5~\mathrm{dBm}\), where PA consumption dominance becomes increasingly pronounced with frequency (Fig.~\ref{fig:TX_Pdc_Model_5dBm_PA_and_0dBm_PA}, bottom). In all cases, the contribution of the mixer to the overall power consumption is marginal. As a result, the lowest aggregate power consumption happens in cases where the mixer yields a high power and a PA is not required. 
These trends underscore the need for the modeling and co-design of the entire transmitter in order to minimize power consumption.


\subsection{Receiver Power Consumption}\label{Section: RX_Modeling}
Receiver power consumption is evaluated by sweeping the ED RF input power from -20 to -5 dBm. Assuming a constant LNA input power of 1 $\upmu$W (-30 dBm), this leads to LNA gains ranging from 10~dB to 25~dB. This analysis is repeated considering two noise figure levels of 7 dB and 15 dB. 

The results, outlined in Fig.~\ref{fig:RX_Pdc_Model_5dBm_PA_and_0dBm_PA}, show different clear trends. First, both the LNA and ED power consumption grow as the RF input power level is increased. In the former, this is due to an increase in the LNA gain to achieve the required RF input power; whereas, in the latter, this is due to the larger currents driving the ED circuits. Second, the LNA power consumption increases sharply with frequency due to its worse FOM. Third, the ED consumption is inversely proportional to frequency, as described in prior sections. Fourth, a more relaxed LNA noise figure (bottom plot) leads to a marked decrease in LNA consumption with no change in the ED power. 
These behaviors reveal an intrinsic trade-off between ED and LNA power across frequency, power levels, and noise figure specs. This affects the transmitter-receiver co-design process and suggests that there may be optimal design regions when balancing the contribution of both sides of the transceiver. 



\subsection{Channel-Aware Transceiver Power Consumption}
\label{Section: TRX_Modeling}

\begin{figure}[!t]
  \centering
  \includegraphics[width=\linewidth]{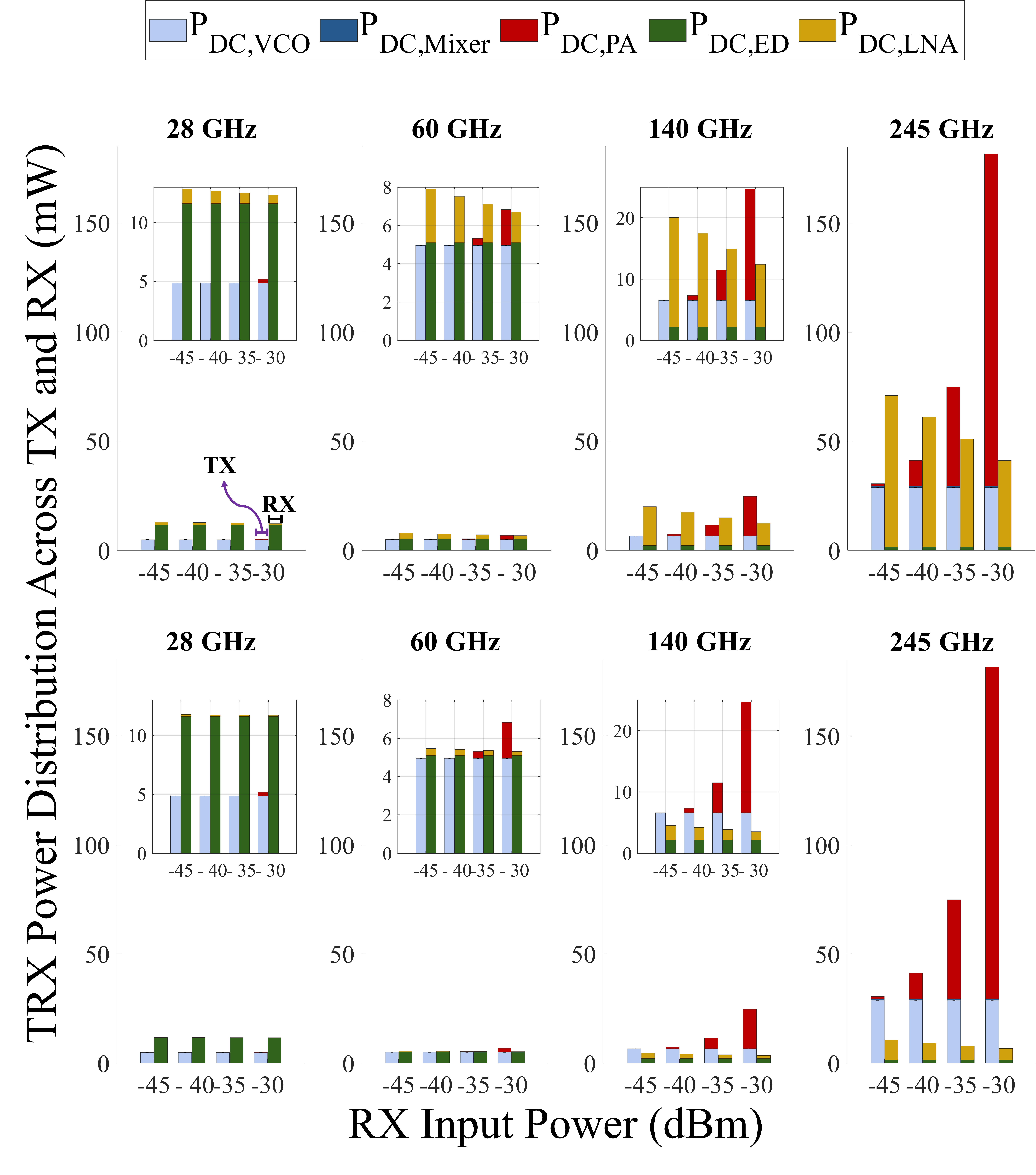}
  \vspace{-0.6cm}
\caption{TRX power breakdown as a function of receiver input power for different carrier frequencies, with fixed $P_{\mathrm{in,PA}} = P_{\mathrm{out,LNA}} = -10$~dBm. Top: $\mathrm{NF}=7$~dB, $R_b=11$~Gbps. Bottom: $\mathrm{NF}=15$~dB, $R_b=2.2$~Gbps.}
\label{fig:TRX_blackbox}
\vspace{-0.2cm}
\end{figure}

As described in Section~\ref{sec:channel}, representative path loss values of 24, 28, 32, and 36 dB are considered for 28, 60, 140, and 245 GHz, respectively. By integrating these channel path loss values into the TX–RX link budget, a channel-aware power consumption model for WNoC systems is derived. Our study aims to show the power consumption of TX and RX blocks, respectively, across frequencies and considering increasing levels of received power at the receiver antenna. We also consider two cases: a first receiver with a noise figure of $NF = 7$ dB, which under the link budget and gains considered lead to a $E_b/N_0$ compatible with $R_b = 11$ Gbps at $BER =$ 10\textsuperscript{-12}, and a second receiver with a relaxed noise figure of  $NF = 15$ dB leading to a slower operation at $R_b = 2.2$ Gbps for the same BER.


Fig.~\ref{fig:TRX_blackbox} shows the results for the TRX power analysis, with the stringent and relaxed NFs at the top and bottom plots respectively. The plots illustrate how the dominant power contributors within the TRX vary with the design and operational conditions. Logically, the RX tends to dominate when the transmitter gain is low and hence the receiver input power is also low; the break-even point, however, depends on frequency. As frequency increases, the transmitter PA consumption increases sharply and becomes more important even at relatively low received power levels. The comparison between top and bottom plots also reveals that, as we relax the noise figure requirements of the receiver, the LNA contribution is sharply decreased. This leaves the transmitter as the most power-hungry component, by far, at high frequencies.



\vspace{-0.2cm}
\section{Discussion}
\label{Section: Discussion2}
The results obtained in the previous section reveal a frequency-dependent TX–RX power trade-off governed by NF and data rate. A lower NF (higher sensitivity) shifts the power burden toward the receiver, increasing overall consumption, whereas relaxed sensitivity requirements favor a shift toward the transmitter. Consequently, achieving energy-efficient WNoC operation requires the joint optimization of transmitter power, receiver gain, and operating frequency rather than isolated subsystem-level tuning, with the catch being that the balance between transmitter and receiver contributions very much depend on the frequency being used, as it determines the channel loss and forces RF sub-blocks to operate closer to the maximum frequencies of the transistors.

\begin{figure}[!t]
  \centering
  \includegraphics[width=\linewidth]{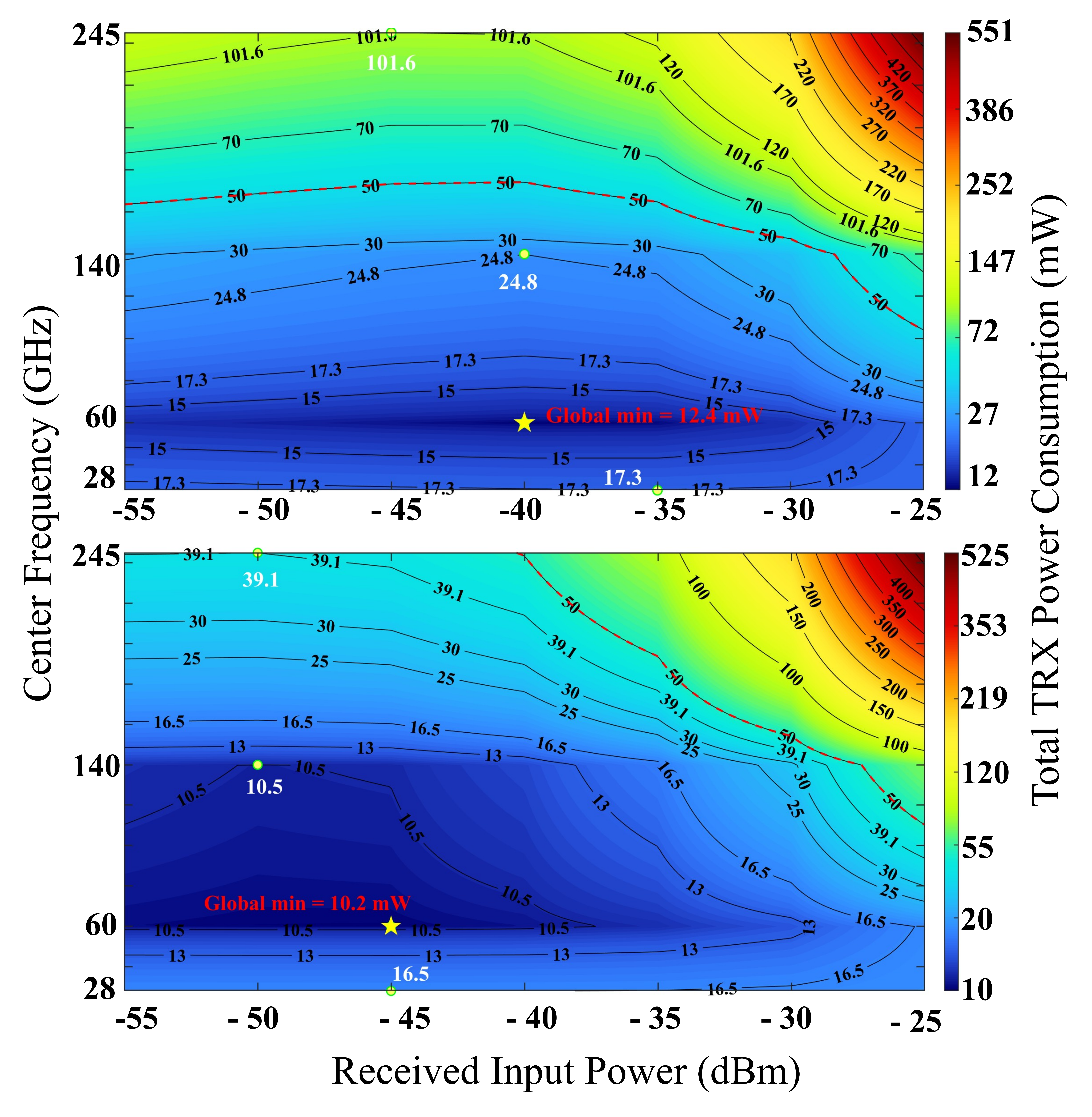}
\caption{Contour plot of TRX DC power vs. received power at different frequencies. Top: NF = 7 dB with $R_{b}$ = 11 Gbps. Bottom: NF = 15 dB with $R_{b}$ = 2.2 Gbps.}
\label{fig:TRX_colormap}
\vspace{-0.5cm}
\end{figure}

We argue that one of the main advantages of our proposed approach is that it allows for quick explorations of the design space to uncover possible optimal regions of performance and guide decisions such as the frequency band to use. In this sense, we show in Fig.~\ref{fig:TRX_colormap} how the prior analysis can be extended by mapping the continuous dependence of total TRX power consumption on both frequency and received input power. Probably because of a combination of technology maturity and bandwidth considerations, the region of optimality seems to be around a frequency of 60 GHz with rather low received input power (that would minimize the TX PA gain). The optimal point, however, remains in an intermediate received input power, underscoring the importance of balancing PA and LNA gains.

Under identical channel assumptions, tightening the NF constraint (Fig.~\ref{fig:TRX_colormap}, top) shifts the optimal operating region toward higher received power levels and increased LNA consumption, thereby raising the overall power floor. Conversely, the relaxed NF scenario (Fig.~\ref{fig:TRX_colormap}, bottom) sees the optimal design point to go not only to lower received input powers, putting more emphasis on the receiver end, but also to higher frequencies. Indeed, at lower frequencies, relaxing the NF results in only a marginal decrease in total power consumption, whereas in the sub-THz regime, a relaxed NF constraint leads to a substantial reduction in required power. In any case, these contour plots provide a compact representation of feasible operating regions and highlight the sensitivity of TRX power to joint sub-block design choices.

Besides providing the global optima, Fig.~\ref{fig:TRX_colormap} also shows the optimal power allocation for the four specific frequencies considered in this work: 28, 60, 140 and 245 GHz. We see how higher frequencies require lower received power to reach optimality, yet the overall power consumption increases. For instance, at 28 GHz with an equivalent receiver NF of 7 dB and a received power of -30 dBm, the TRX consumes 17.3~mW. In contrast, at sub-THz frequencies (245 GHz), the power consumption scales to 101.6 mW, even with a lower received power of -40 dBm and the same 7 dB NF. 


Another interesting tradeoff to explore is the balance between power and speed. Relaxed noise figures reduce power consumption, but also raises the effective noise floor, reducing the allowable SNR margin and thus the maximum achievable data rate for a fixed BER. In this context, we plot the bit-energy efficiency $E_{\mathrm{bit}} = P_{DC,TRX}/R_{b}$ in pJ/bit as a function of the receiver noise figure and, hence, the assumed data rate for a fixed BER. The plots are obtained assuming a fixed received input power equal to the minimum required level to satisfy the target BER, hence assuming a common data rate across all operating frequencies for each value of NF. 

\begin{figure}[!t]
  \centering
  \includegraphics[width=\linewidth]{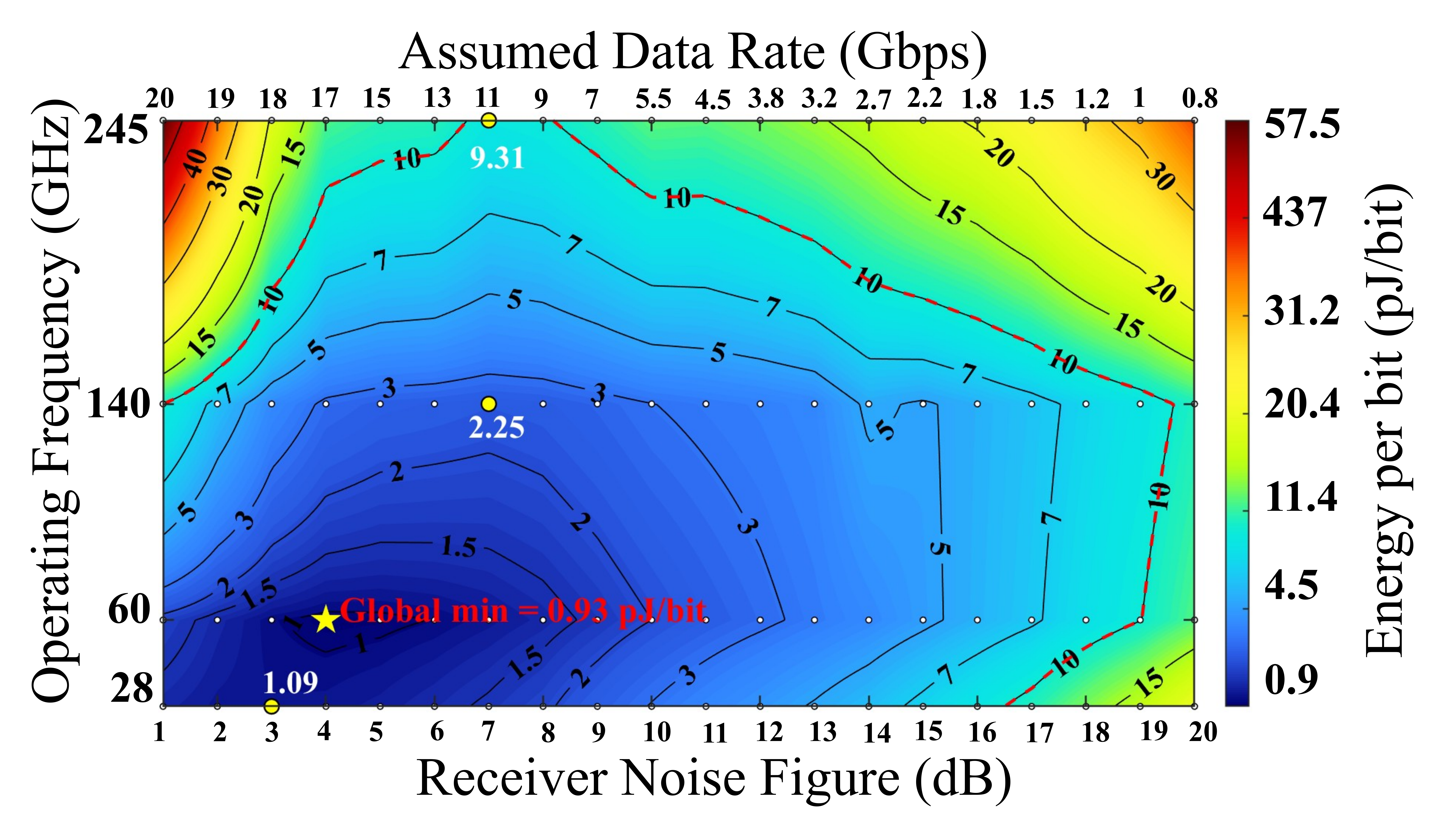}
  \vspace{-0.4cm}
\caption{Contour plot of TRX energy per bit versus carrier frequency and receiver NF, highlighting optimal regions and global minimum. For each NF, the received power is fixed to the NF-dependent optimal value and applied across all frequencies. The red dashed line indicates the 10 pJ/bit contour.}\label{fig:TRX_pJ_per_bit_colormap}
\vspace{-0.4cm}
 \end{figure}

The resulting energy efficiency landscape, shown in Fig.~\ref{fig:TRX_pJ_per_bit_colormap}, exhibits a non-monotonic behavior and reveals an optimal operating region rather than a simple trade-off with NF or data rate alone. In particular, the global minimum occurs at 60~GHz with NF = 4~dB, achieving an energy efficiency of approximately 0.93~pJ/bit at a data rate of 17~Gbps. The reason for such an optimal point can be explained because (1) beyond 60 GHz, the energy per bit required increases with the frequency of the transceiver, and because (2) relaxing the NF and, hence, reducing the transmission speed, also tends to lead to an increase in the energy per bit. The former trend is due to the more inefficient sub-systems and higher channel loss at higher frequencies, which counterbalance the potential bandwidth advantage; whereas the latter trend is due to the fact that the power of the TRX blocks is better amortized at higher speeds, even if the required LNA consumes higher power.

It is worth noting that the current model and the results obtained in the exploration are a reflection of the current state of the art in the high-frequency RF design field, rather than a prediction of future performance. Therefore, as technology and RF circuit design continues to mature, we may see a modification of the power consumption landscape perhaps shifting the optimal points slightly to higher frequencies. Yet still, due to $f_{T}$ and $f_{max}$ restrictions of CMOS technology, one can expect higher frequencies to be inherently less efficient, despite evident bandwidth advantages. 



\vspace{-0.1cm}

\section{Conclusion}\label{Section: Conclusion}
This work presents a data-driven, channel-aware behavioral model for estimating the power consumption of CMOS OOK TRXs in WNoC systems. The framework explicitly incorporates frequency-dependent channel path loss into a unified TRX representation and enables per-sub-block power estimation (PA, VCO, mixer, LNA, and ED), while capturing their joint impact on overall system behavior. The proposed methodology allows to explore the frequency-dependent trade-offs between the power consumed by the transmitter and the receiver, enabling their co-design in multiple scenarios. Applying the methodology to the exploration of WNoC systems has revealed the existence of frequency-dependent sweet spots and a global minimum in the energy-per-bit landscape. In particular, it shows that an energy consumption of less than 1 pJ/bit in CMOS is achievable by working at 60~GHz and operating close to 20 Gbps. We finally note that the proposed method can be extended to other technologies, modulation schemes, or application spaces, and that it can be even connected to design space exploration and optimization frameworks to find global optima of power consumption over arbitrary channel models across a broad range of frequencies.





\vspace{-0.2cm}

\section*{Acknowledgment}

The authors acknowledge support from the European Commission through grants HORIZON EIC 2022 PATHFINDEROPEN 01-101099697 (QUADRATURE), HORIZON ERC 2021-101042080 (WINC), and HORIZON MSCA 2022-PF-101062272 (ENSPEC6G), and from TUBITAK through grant 121C073.

\bibliographystyle{IEEEtran}
\bibliography{References.bib}
\vspace{-35pt}
\begin{IEEEbiography}[{\includegraphics[width=1in,height=1.1in,clip,keepaspectratio]{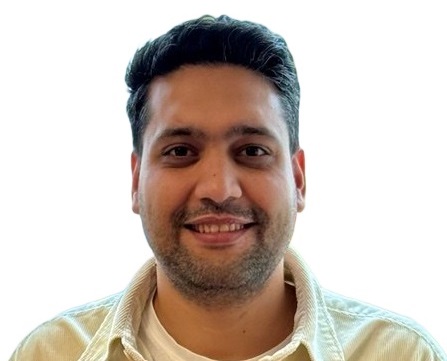}}]{Mohammad Shahmoradi}{\space}(Student Member, IEEE) received the B.Sc. and M.Sc. degrees in electrical engineering from Shahrekord University in 2018 and from the Iran University of Science and Technology (IUST), Tehran, Iran, in 2021, respectively. Since September 2023, he has been pursuing the Ph.D. degree at the Universitat Politècnica de Catalunya (UPC), Barcelona, Spain, funded by a European Research Council scholarship. His research interests encompass energy-efficient wireless Network-on-Chip architectures, the design of broadband RF and microwave components, including power amplifiers, as well as the study of mm-wave and sub-THz CMOS circuits.
\end{IEEEbiography}\vspace{-35pt}
\begin{IEEEbiography}[{\includegraphics[width=1in,height=1.1in,clip,keepaspectratio]{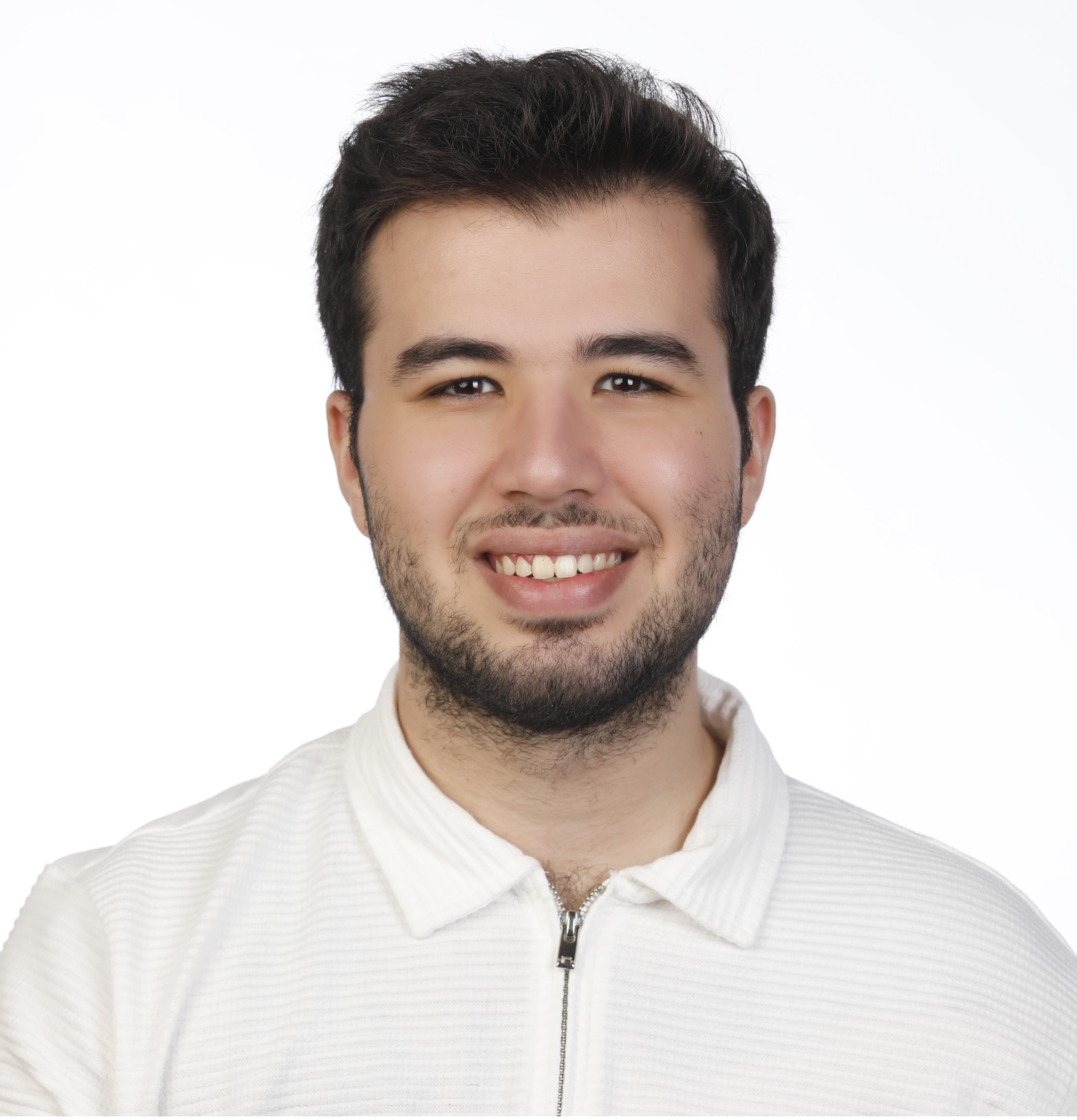}}]{Ahmet Yelboga}{\space}(Graduate Student Member, IEEE) received the B.S. degree in Electronics Engineering from Sabanci University in 2025, where he is currently pursuing the M.S. degree. His research interests include millimeter-wave and sub-terahertz CMOS circuits, focusing on energy-efficient designs such as modulators and non-coherent demodulators, as well as broadband and distributed CMOS circuits, including distributed amplifiers and distributed mixers. He is a recipient of the 2025 IEEE MTT-S Undergraduate/Pre-Graduate Scholarship Award for his research on an energy-efficient CMOS OOK transmitter.
\end{IEEEbiography}\vspace{-35pt}
\begin{IEEEbiography}[{\includegraphics[width=1in,height=1.1in,clip,keepaspectratio]{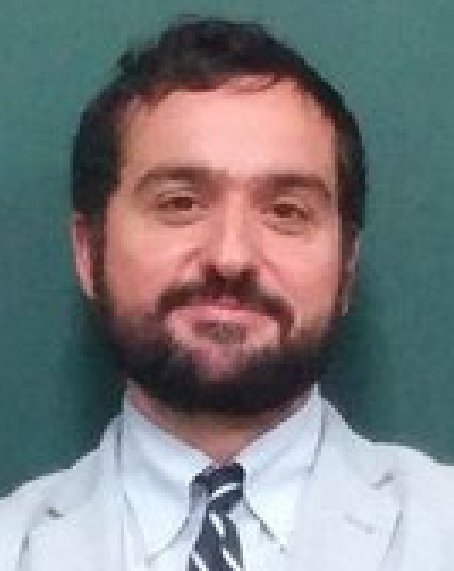}}]{Eduard Alarcón}{\space}(Member, IEEE) received the M.Sc. and Ph.D. degrees in EE from UPC BarcelonaTech, Spain, in 1995 and 2000, respectively. He is a Full Professor with the School of Telecommunications, UPC. He was a Visiting Professor with the University of Colorado at Boulder, USA, and the Royal Institute of Technology (KTH), Stockholm. He has been involved in different European (H2020 FET-Open and FlagERA) and U.S. (DARPA and NSF) research and development projects within his research interests, including the areas of on-chip energy management, wireless networks-on-chip, machine learning accelerator architectures, and nanotechnology-enabled wireless communications. He has received the Google Faculty Research Award in 2013; the Samsung Advanced Institute of Technology Global Research Program Gift, in 2012; and the Intel Honor Program Fellowship in 2014. He received the National Award for master’s study. He was the Vice President of Technical Activities of IEEE CAS (2016–2017 and 2017–2018). He has organized and chaired the conferences, including the General Co-Chair of 2020 IEEE International Symposium on Circuits and Systems 2020, Seville, Spain; the Special Sessions Co-Chair of 2013 IEEE International Symposium on Circuits and Systems, Beijing, China; the Special Sessions Chair of IEEE ISCAS 2023 Monterrey, CA, USA, the TPC Co-Chair of IEEE ISCAS 2025, Shanghai; and a Steering Committee Member of IEEE AICAS and IEEE IBM-CAS Symposium devoted to AI chip architectures (2017–2020). He has chaired special sessions on quantum computers design in DATE and IEEE QCE 2023 and QCE2024. From 2018 to 2019, he served as the Editor-in Chief for IEEE JOURNAL ON EMERGING TOPICS IN CIRCUITS AND SYSTEMS.
\end{IEEEbiography}\vspace{-35pt}
\begin{IEEEbiography}[{\includegraphics[width=1in,height=1.1in,clip,keepaspectratio]{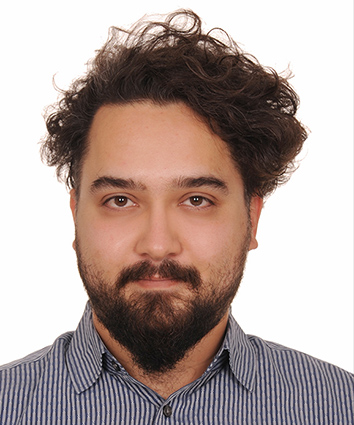}}]{Korkut Kaan Tokgöz}{\space} (Member, IEEE) received the B.Sc. and M.Sc. degrees from the Electrical and Electronics Engineering Department, Middle East Technical University, Ankara, Türkiye, in 2009 and 2012, respectively, and M.Eng. and Ph.D. degrees from the Department of Physical Electronics, Tokyo Institute of Technology, Tokyo, Japan, in 2014 and 2018, respectively. From 2018 to 2019, he was a Senior Researcher/an Assistant Manager with NEC Corporation, Kanagawa, Japan, where he was involved in 5G systems and fixed point-to-point wireless links. From 2019 to 2022, he was an Assistant Professor with Tokyo Institute of Technology. He is currently a Faculty Member with the Faculty of Engineering and Natural Sciences, Sabanci University, Istanbul, Türkiye. He is Co-Founder and CTO of Evrim Company Ltd., Yokohama, Japan. He is a founding member of the Turkish Integrated Circuits Alliance (TICA) and serves as IEEE SSCS YP Chair. His research interests include analog/RF/millimeter-wave/sub-terahertz TRXs, low-power Edge-AI for monitoring systems, IoT, sensors, de-embedding, device characterization, high-power and high-efficiency PAs for wireless systems. He was a recipient of several awards, scholarships, and grants, including the TUBITAK 2232-B International Fellowship for Early-Stage Researchers 2022, the Marie Skłodowska-Curie Actions Post-Doctoral Fellowship 2022, the SSCS Predoctoral Achievement Award, in 2018, the IEEE MTT-S Graduate Student Fellowship, in 2017, the IEICE Student Encouragement Prize, in 2017, the Seiichi Tejima Overseas Student Research Award, and the IEEE/ACM ASP-DAC University LSI Design Contest 2017 Best Design Award.
\end{IEEEbiography}\vspace{-35pt}
\begin{IEEEbiography}[{\includegraphics[width=1in,height=1.1in,clip,keepaspectratio]{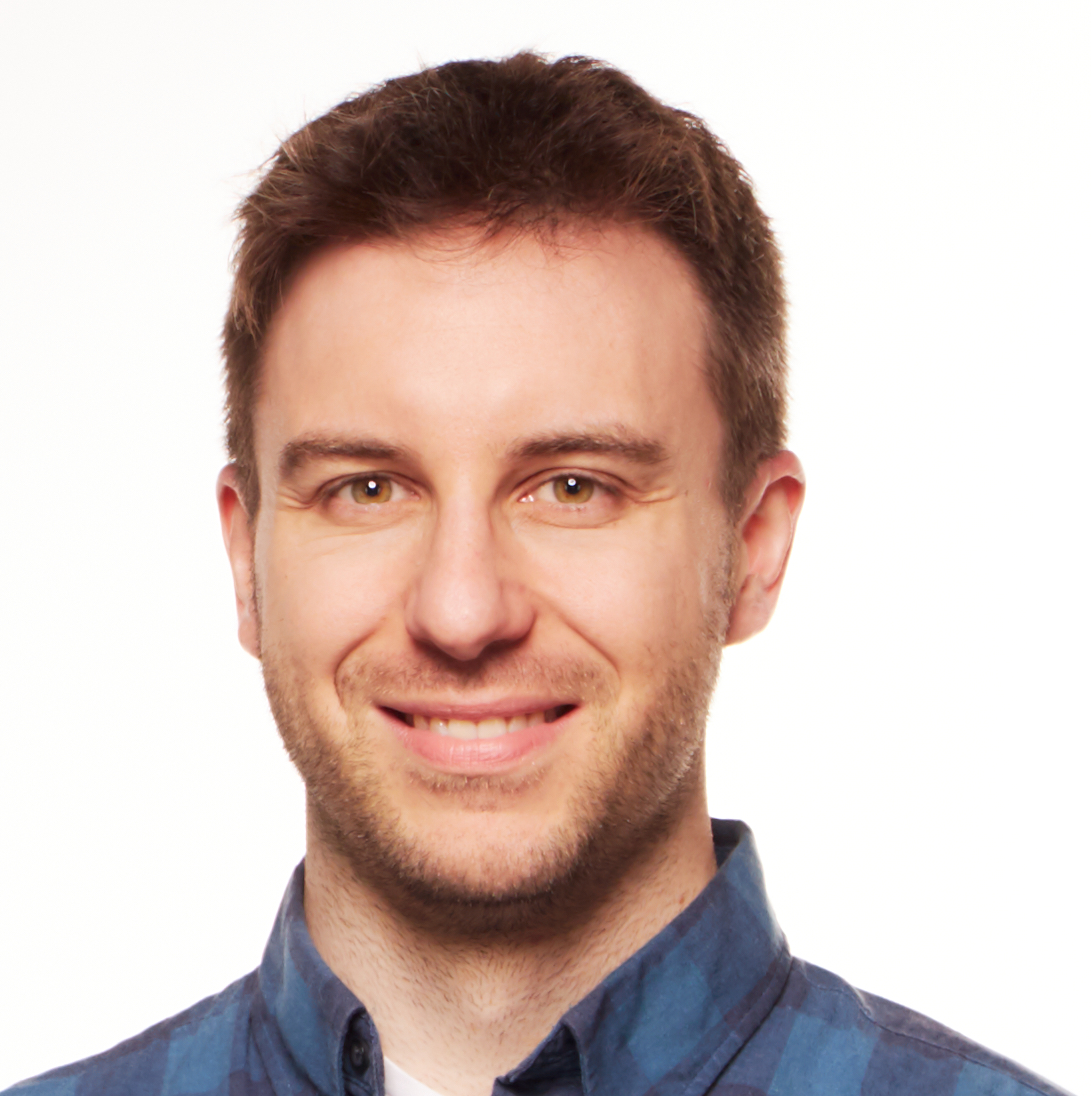}}]{Sergi Abadal}{\space}(Senior Member, IEEE) is an Associate Professor at the Universitat Politècnica de Catalunya (UPC) and Director of the Nanonetworking Center in Catalunya (N3Cat). He is the recipient of an ERC Starting Grant and also is or has been the coordinator of multiple EU projects. He holds editorial positions in journals such as the IEEE TMC, IEEE TCAD, or IEEE JETCAS. He has served as a TPC member of more than 40 conferences and has published over 150 articles in top-tier journals and conferences. Thanks to his efforts, in 2025, he received the \emph{Matilde Ucelay} national research award from the Ministry of Science as the best young researcher in Spain working in engineering. He also received the \emph{Agustín de Betancourt} medal awarded by the Spanish Royal Academy of Engineering to the top 10 young researchers in Spain in 2024, the ACM NanoCom Outstanding Milestone Award in 2022 and the Young Investigator Award of the Nano Communication Networks Journal in 2019. His current research interests are in the areas of wireless communications in extreme environments and applications.
\end{IEEEbiography}

\end{document}